\documentclass[amsmath,amssymb,twocolumn,aps,prb]{revtex4-2}
\usepackage{graphicx}
\usepackage{epstopdf}
\usepackage{bm}
\usepackage{subfigure}
\usepackage{sidecap}

\usepackage{graphicx}
\usepackage{color}
\usepackage{epstopdf}
\usepackage{amssymb}
\usepackage{amsmath}

\usepackage{bm}
\graphicspath{{Figures/}}
\usepackage{subfigure}
\usepackage{sidecap}
\usepackage {times}

\newcommand{\si}{\sigma}

\newcommand{\ga}{\gamma}
\newcommand{\eps}{\epsilon}
\newcommand{\vare}{\varepsilon}

\newcommand{\lam}{\lambda}

\newcommand{\de}{\delta}
\newcommand{\De}{\Delta}

\newcommand{\be}{\begin{equation}}
\newcommand{\ee}{\end{equation}}

\newcommand{\bea}{\begin{eqnarray}}
\newcommand{\eea}{\end{eqnarray}}
\newcommand{\bd}{\begin{displaymath}}
\newcommand{\ed}{\end{displaymath}}
\newcommand{\ba}{\begin{array}}
\newcommand{\ea}{\end{array}}
\newcommand{\bi}{\begin{itemize}}
\newcommand{\ei}{\end{itemize}}
\newcommand{\bc}{\begin{center}}
\newcommand{\ec}{\end{center}}
\newcommand{\bfl}{\begin{flushleft}}
\newcommand{\efl}{\end{flushleft}}
\newcommand{\bfr}{\begin{flushright}}
\newcommand{\efr}{\end{flushright}}
\newcommand{\non}{\nonumber}

\newcommand{\bl}{\begin{aligned}}
\newcommand{\el}{\end{aligned}}

\newcommand{\hf}{\hat{f}}

\newcommand{\hT}{\hat{T}}

\newcommand{\hh}{\hat{h}}

\newcommand{\hDe}{\hat{\Delta}}

\newcommand{\home}{\hat{\omega}}

\newcommand{\blam}{\bar{\lambda}}
\newcommand{\bn}{\bar{n}}

\newcommand{\fs}{\frac{1}{2}}

\newcommand{\om}{i\omega_n}

\newcommand{\ra}{\rangle}
\newcommand{\la}{\langle}

\newcommand{\PR}{PrIr$_3$}

\def\bR{{\bf R}} 
  \def\bq{{\bf q}}

\def\bQ{{\bf Q}}   
  
 \def\bd{{\bf d}}

\def\={\!\!\!&=&\!\!\!}
\def\+{\!\!\!&&\!\!\!+~}
\def\-{\!\!\!&&\!\!\!-~}

\usepackage{color}
\usepackage[dvipsnames]{xcolor}
\usepackage{xcolor}
\def\BLU{\color{black}}  %

\usepackage[colorlinks=true,citecolor=blue]{hyperref}

\usepackage{hyperref,cleveref}

\begin{document}

\title{Magnetocalorics of singlet ground state induced moment magnets}

\author{Peter Thalmeier}
\affiliation{Max Planck Institute for Chemical Physics of Solids, D-01187 Dresden, Germany}
\date{\today}

\begin{abstract}
In f-electron materials  like Pr or U compounds with non-Kramers states of the f-shell the ground
state may be a nonmagnetic singlet due to the action of the crystalline electric field. Nevertheless these
compounds can order magnetically. They develop a ground state moment and long range order due
to the spontaneous admixture of the excited state into the ground state caused by inter-site exchange.
This mechanism can establish magnetic order  if a control parameter 
exceeds a critical value defining the quantum critical point between disordered and magnetic phase.
Here we investigate the magnetocaloric properties of such quantum magnets where the entropy
release at low temperature is due to a competition between thermal depopulation and spontaneous order effects.
We determine field and temperature dependence of specific heat for ferro- and antiferromagnetic order
and also calculate the adiabatic magneto- and barocaloric cooling rates. As a model application we discuss 
the magnetic specific heat of the new induced ferromagnet PrIr$_3$.
Furthermore we analyze the excitation spectrum of the singlet induced moment magnets in an external 
field with particular emphasis on field and temperature dependence of the critical soft mode. We find that 
the latter is fragile and exists only at specific points in the phase diagram.

\end{abstract}
\maketitle

\section{Introduction}
\label{sec:intro}

There is a class of genuine quantum magnets that cannot be described by the common semiclassical
picture\cite{majlis:07,schmidt:17} of spontaneous order of preformed  moments from degenerate ground state multiplets. Most frequently these are  induced moment quantum magnets with localized {\it singlet} ground states and another close by singlet or doublet, triplet state (depending on the site symmetry). The splitting is caused by the effect of the crystalline electric field (CEF) on the total angular momentum J-multiplets of f-electron shells. Examples of such metallic f-electron materials will be discussed in later Sections.
The nonmagnetic ground state may 
nevertheless acquire a collective ordered moment by spontaneous mixing of the excited state into the ground state through the effective {\BLU RKKY inter-site exchange which is mediated by  conduction electrons. For a singlet ground state the spontaneous appearance of an ordered moment requires the effective exchange to exceed a critical limit as compared to the splitting
energy of ground and excited states.  Increasing this control parameter beyond a critical value that defines the quantum critical point (QCP), a transition from paramagnetic to a ground state with induced magnetic order occurs}. This mechanism is also possible for induced order of higher order multipolar moments like (ferro-)quadrupolar order in 4f materials. \\

The various possible f-electron type induced moment order has been observed in numerous Pr- and U- compounds. For a comparative investigation and examples see Ref.~\cite{thalmeier:24} and the references cited therein. In the present work we extend the analysis of singlet-singlet induced quantum magnetism under the application of an external field, considering ferro- (FM) as well as antiferro- (AFM) magnetic cases. {\BLU We will present comprehensive results firstly on the temperature- and field behavior of the induced (sublattice-) moments, the associated H-T phase boundary as well as the critical field as function of control parameter. Secondly particular emphasis is placed on a new systematic investigation of  magnetocaloric properties of the singlet-singlet induced moment magnets. 
We calculate field dependence of specific heat  jump anomalies in relation to the underlying Schottky behavior for FM as well as AFM order and demonstrate a strikingly distinct behaviour between the two cases. For AFM the relative position of the jump to the Schottky maximum position gives a measure to estimate the control parameter. Furthermore we show that the low temperature susceptibility exhibits a pronounced field dependence for control parameters close to the QCP. These predicted
thermodynamic properties can be useful to characterise singlet ground state materials and as an example a preliminary application to the recently discovered induced moment ferromagnet \PR~\cite{gornicka:24} is made. 
We also compare the two different types of competing entropy release in the induced moment magnets which is not present
in the semiclassical degenerate ground state magnets. As another new investigation in these systems we calculate the adiabatic magnetic cooling rate. As for the specific heat the cooling rates are very different for FM and AFM cases. In particular in the former
it will be strongly enhanced for low fields and low temperature when the control parameter is slightly above the critical one.
This effect should be  of considerable experimental interest.
Furthermore we consider the analogous  barocaloric cooling rate caused by the Gr\"uneisen parameter coupling to the CEF  splitting of singlets which therefore tunes the control parameter. A large step-like anomaly at the critical pressure is predicted showing a different field dependence for FM and AFM cases.

Finally the dynamical properties of the singlet-singlet FM and AFM are studied by deriving the temperature- and field dependence of magnetic exciton dispersion. We find that the soft mode behaviour in the AFM around the field dependent transition temperature is fragile and unexpectedly  becomes arrested between zero and critical field. This new prediction may be tested by inelastic neutron scattering in singlet ground state AFM.}\\

The ingredients of the two singlet model are defined in Sec.~\ref{sec:singmod}. Its molecular field (MF) treatment in Sec.~\ref{sec:MFA} leads to the selfconsistent solutions for the order parameters and renormalised singlet splitting in the ordered phase as well as critical temperature and field dependence on the control parameter. In the main Sec.~
\ref{sec:caloric} we derive thermodynamic potentials, specific heat and adiabatic caloric coefficients which are calculated numerically using the MF solutions. Sec.~\ref{sec:materials} {\BLU gives an overview on f-electron singlet ground state materials and in particular a brief discussion of the new induced ferromagnet \PR~in the context of the two singlet model.} Finally in Sec.~\ref{sec:magex} we analyze in detail the field- and temperature dependent magnetic exciton dynamics and their soft mode characteristics. A discussion of numerical results and a summary are presented in Secs.~\ref{sec:discussion},\ref{sec:summary}.

\section{The singlet-singlet CEF model in an external field}
\label{sec:singmod}

{\BLU The localized states of 4f and 5f electron shells can approximately be characterized by a total angular momentum (J) multiplet.
The action of the crystalline electric field (CEF) splits it into a sequence of CEF sub- multiplets depending on the site symmetry of
the f element \cite{lea:62,hutchings:64}. For non-Kramers ions (integer J=4,6,8) it is possible that the lowest energy states are two singlets. This  can occur generally
in uniaxial or lower site symmetries \cite{thalmeier:21,thalmeier:24}. A list of material examples is given in Sec.~\ref{sec:materials}.
A lot of qualitative understanding of induced singlet ground state magnetism may be derived from
this simple singlet-singlet CEF model. As mentioned in the introduction it has been investigated in particular in context with  (J=4) $4f^2$ Pr and $5f^2$ U intermetallic compounds  in sufficiently low site symmetry.}  Without field the thermodynamic, quantum critical and dynamical properties of the two singlet model have also been studied in detail in connection with other common singlet-doublet or -triplet CEF systems \cite{thalmeier:24}. The two singlet model in pseudo-spin representation may be written as
\bea
H=\De\sum_iS_{zi}-m_s^2I_0\fs\sum_{\la ij\ra}S_{xi}S_{x_j}-m_sh\sum_iS_{xi},
\label{eq:Ham}
\eea
where the pseudo spin states with $S_z=\pm\fs$ correspond to the CEF ground state $|0\ra\equiv |-\fs\ra$ and 
$|1\ra\equiv |+\fs\ra$ excited state at energies $-\frac{\De}{2}$ and $+\frac{\De}{2}$, respectively. Here 
$m_s=\la 0|J_z|1\ra$ is the matrix element of the physical 4f-shell total angular momentum $J_z$ between the two CEF states. In the $(|0\ra, |1\ra)$ subspace $J_{x,y}$ have not matrix elements. In the pseudo spin representation
$S_x$ takes over the role of $J_z$, i.e. $J_z\rightarrow m_sS_x$ and likewise $S_{y}$ does not appear in the above Hamiltonian. Furthermore $h=g_J\mu_BH$ where $g_J$ is the Land\'e factor and H the strength of the external field oriented along physical (uniaxial) z-direction. The effective exchange interaction is given by $I_e=z|I_0|$ where $I_0$ is the nearest-neighbor (n.n.) exchange constant and z the coordination number. We use the convention $I_0 > 0$ for ferromagnetic (FM) exchange and $I_0<0$ for antiferromagnetic (AFM) exchange.

We consider two possible FM or AFM broken symmetry ground states. 
In MF approximation the model described by Eq.~(\ref{eq:Ham}) reduces to the sum $\sum_i h_{mf}(i)$ over single-site Hamiltonians. In unified notation for the two cases ($\sigma=+1$ for FM and $\sigma=-1$ for AFM) the single site terms are given by
\bea
h^\lam_{mf}(i)&=&\hh^\lam_{mf}(i)+\fs\sigma m_s^2 I_e \la S_x\ra_\lam \la S_x\ra_{\blam}\non\\
\hh^\lam_{mf}(i)&=&\De S_z(i)-m_sh_e^\lam S_x(i)\non\\
&=&\frac{\De}{2}\left(
 \begin{array}{cc}
1& -\ga'_\lam\\
-\ga'_\lam&-1\\
\end{array}
\right),
\label{eq:MFHam}
\eea
with h denoting the external and $h_e^\lam$ the effective field. Here  $\lambda =A,B$ (with corresponding $(\blam=B,A$) denotes one of the two sublattices in the AFM case. In the FM case they are identical, i.e.  $\la S_x\ra_\lam \equiv \la S_x\ra$. We note that in the present model the moment is frozen along field direction by the CEF splitting $\De$. Therefore not only the sign of $\la S_x\ra_{A,B}$ may be opposite but their magnitude is also different for nonzero applied field. Explicitly the effective molecular field $h^\lam_e$ or its dimensionless form $\gamma'_\lam$ is defined by
\bea
h_e^\lam&=&h+\sigma m_sI_e\la S_x\ra_{\blam}\non\\
\ga'_\lam&=&\frac{m_sh_e^\lam}{\De}=m_s(h'+\si\ga\la S_x\ra_{\blam}),
\label{eq:heff}
\eea
where $h'=h/\De=g_J\mu_BH/\De$ and $\gamma=m_sI_e/\De=(2/m_s)\xi_s$ are dimensionless external field and interaction parameter. Here the important control parameter for the quantum phase transition from nonmagnetic to magnetic ground state of the model is introduced by
%
\begin{figure}
\hspace{-0.5cm}
\includegraphics[width=0.75\columnwidth]{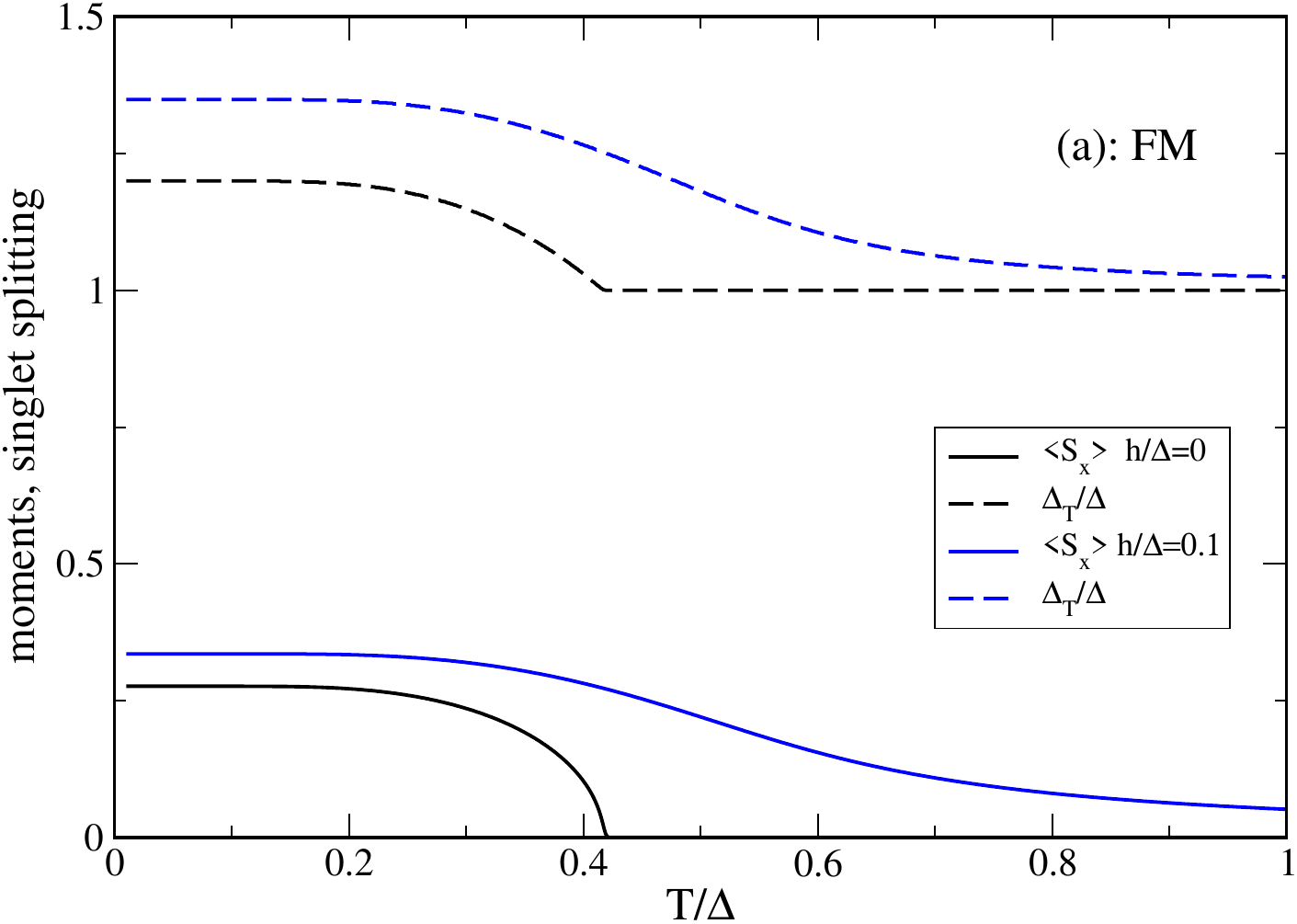}\\
\includegraphics[width=0.90\columnwidth]{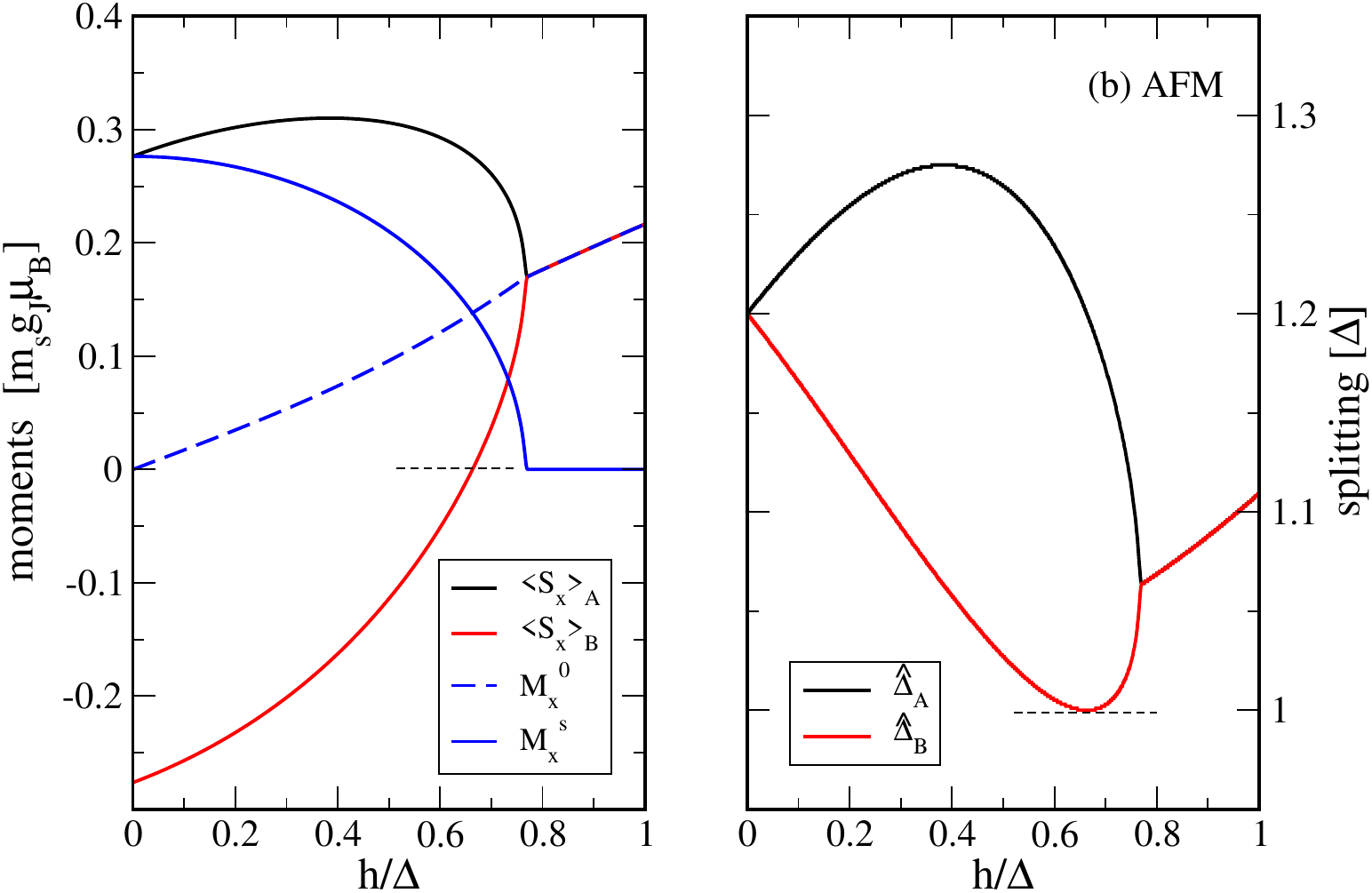}
\caption{(a) Homogeneous moment $\la S_x\ra=M_x^0$ as function of temperature for $\xi_s=1.2$. 
At zero field the ordering temperature is $T_m/\De=0.42$. At finite field the FM phase transition changes
into a gradual crossover at $T_m$. Dashed lines show the renormalised (by order parameter and field)  singlet-singlet 
splitting $\hDe_T=\De_T/\De$.
(b) Left panel: AFM sublattice (A,B) moments, homogeneous $(M_x^0)$ and staggered  $(M_x^s)$ moments 
as function of applied field $h'=h/\De$ at $T/\De=0.01$  for a control parameter $\xi_s=1.2$ leading to critical field $h'_c=0.77$. The induced AF phase transition persists for all fields up to $h'_c$. Right panel: renormalised A,B sublattice
singlet splittings. The minimum $\hDe_B=1$ appears when $\la S_x\ra_B$ changes sign (black dashed lines).}
\label{fig:FM-AFOP}
\end{figure}
%
\bea
\xi_s=\frac{m_s^2I_e}{2\De}.
\eea
For $\xi_s<1$ or $\xi_s>1$ the ground state will be nonmagnetic or exhibits induced magnetic (FM or AFM) order, respectively \cite{thalmeier:24} with $\xi^c_s=1$ designating the quantum critical point.
{\BLU As a combination of various microscopic parameters of the strongly correlated f-states and their effective intersite
interaction this central important quantity has to be considered as an adjustable model parameter. As we argue later it can be estimated
from specific heat analysis or the dispersion characteristics of magnetic exciton modes. In some cases a tuning of 
the control parameter by applying external (uniaxial) pressure \cite{jensen:87} or substitution with nonmagnetic elements \cite{cooper:70}
has been achieved.}

For the dimensionless field  we may alternatively define $\hh =m_sh'=g_{\text{eff}}\mu_BH/\De$ with the effective g-factor of the two singlets given by $g_{\text{eff}}=m_sg_J$.
{\BLU The above Hamiltonian $\hh^\lam_{mf}$}
is readily diagonalised leading to energy levels
\bea
\eps^\pm_\lam=\pm\eps_\lam=\pm\frac{\De}{2}[1+(\ga'_\lam)^2]^\fs,
\eea
and sublattice $\lam$ =A,B- dependent eigenstates
\bea
|\eps^+_\lam\ra&=&\cos\theta_\lam|1\ra-\sin\theta_\lam|0\ra\non\\
|\eps^-_\lam\ra&=&\sin\theta_\lam|1\ra+\cos\theta_\lam|0\ra,
\label{eq:eigenstate}
\eea
with the mixing angle given by $\tan 2\theta_\lam=\ga'_\lam$.
{\BLU The constant energy shift contained in the first line of Eq.~(\ref{eq:MFHam}) is irrelevant for
calculating the molecular fields but has to be included later for the internal
and free energies in Sec.~\ref{sec:thermpot}.}

\section{Order parameter, magnetisations and critical temperature}
\label{sec:MFA}

The eigenstates of Eq.~(\ref{eq:eigenstate}) may be used to derive
the selfconsistency equations for the mean field order parameter according to
\bea
\la S_x\ra_\lam=\sum_{n=\pm}p^n_\lam\la \eps^n_\lam |S_x| \eps^n_\lam\ra=
\fs\tanh(\beta\eps_\lam)\sin 2\theta_\lam,
\eea
with $p^n_\lam=Z^{-1}_\lam\exp(\beta\eps^n_\lam)$ $(\beta=1/T; k_B\equiv 1)$ denoting the
thermal occupation of the two eigenstates and $Z_\lam=2\cosh(\beta\eps^n_\lam)$ giving
the partition function of the two levels.
Explicitly, using Eq.(\ref{eq:heff}) the order parameter is then obtained from
the selfconsistent equation
\bea
\la S_x\ra_\lam&=&\fs m_s(h'+\si\ga\la S_x\ra_{\blam})\frac{1}{\hDe_T^{\lam}}
\tanh\bigl[(\frac{\De}{2T})\hDe^\lam_T\bigr]\non\\
\hDe^\lam_T&\equiv&\frac{1}{\De}(\eps^\lam_+ -\eps^\lam_-)=[1+m_s^2(h'+\si\ga\la S_x\ra_{\blam})^2]^\fs\non\\
&=&[1+(m_sh'+2\si\xi_s\la S_x\ra_{\blam})^2]^\fs,
\label{eq:MF}
\eea
%
%
\begin{figure}
\includegraphics[width=0.80\columnwidth]{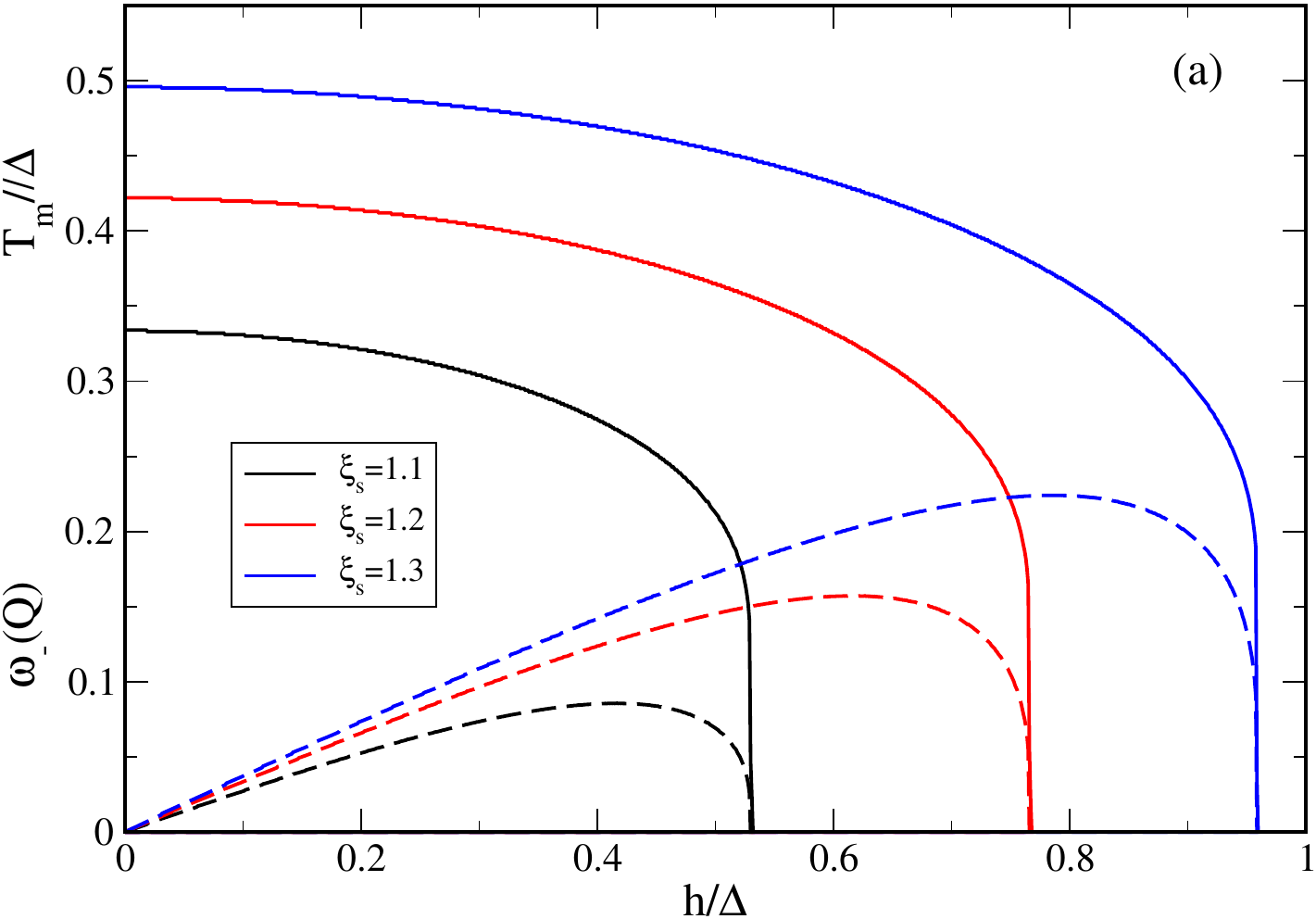}\\
\includegraphics[width=0.80\columnwidth]{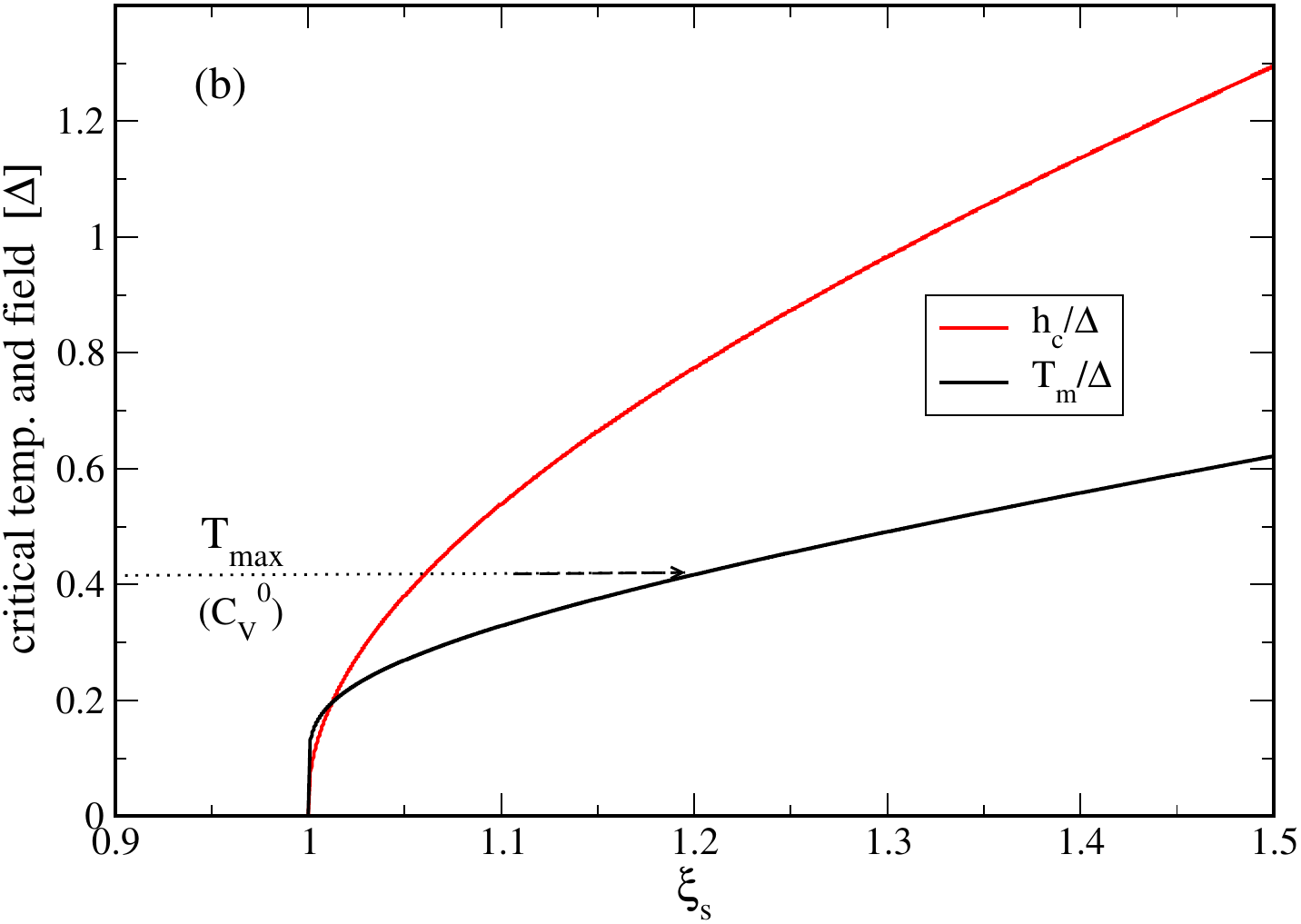}
\caption{(a) Field dependence of AFM induced moment transition temperature (phase diagram) for three different 
control parameters showing the increase of $T_m$ and $h_c$ with $\xi_s$  (full lines). For comparison
the arrested soft mode frequency $\omega_-(\bQ,T_m(h'))$ is shown which is nonzero except at zero and 
critical field (dashed lines). (b) Dependence of AFM critical temperature {\BLU T$_m$ (for h=0) and critical field 
h$_c$ (for T=0)  on the control parameter above the QCP.}}
\label{fig:AF-Tm}
\end{figure}
%

where $\hDe^\lam_T$ is the T-dependent splitting (normalised to $\De$) 
of the eigenstates on each sublattice modified due to the action of the external and molecular fields
resulting from the other sublattice (for the AFM case). Naturally for FM case there is
no distinction between them and the $\lam$ index is irrelevant.  In units of $(g_{\text{eff}}\mu_B/\text{site})$ the dimensionless homogeneous  ($M^0_x$)  and staggered ($M^s_x$)   magnetisation are then given by 
\bea
M^{0,s}_x=\fs[\la S_x\ra_A\pm\la S_x\ra_B],
\label{eq:mag}
\eea
respectively. For the FM and AFM paramagnetic $(h'>h'_c)$ cases there is no sublattice dependence and we have $M_x^0=\la S_x\ra$ and $M_x^s=0$ where the homogenous magnetisation at zero temperature  is now obtained from Eq.~(\ref{eq:MF}) as solution of
\bea
\la S_x\ra_{T=0}=M_x^0=\frac{\fs(m_sh'+2\xi_s M_x^0)}{[1+(m_sh'+2\xi_s M_x^0)^2]^\fs}.
\eea
Here the denominator is the renormalised saturated level splitting $\hDe_0$.
For the calculation of thermodynamic properties it will also be useful to write the selfconsistency equations in a
more compact form. We will use the following identities and abbreviations:
\bea
\ga^{'2}_\lam=(\hDe^\lam_T)^2-1;\;\;\; f_s^\lam(T)=\tanh\bigl[(\frac{\De}{2T})\hDe^\lam_T\bigr].
\label{eq:auxlam}
\eea
The auxiliary function $f_s^\lam(T)$ describes the difference of thermal populations of the molecular field eigenstates.
Then Eq.~(\ref{eq:MF}) may also be written as 
\bea
\la S_x\ra_\lam =sign(\la S_x\ra_\lam)\fs\frac{[(\hDe^\lam_T)^2-1]^\fs}{\hDe^\lam_T} f_s^\lam(T),
\eea
which establishes the direct relation between the order parameters $\la S_x\ra_\lam$ and the (normalized) effective 
splittings $\hDe^\lam_T$ for each sublattice $\lam$ =A,B which are generally different in the AFM
case inside the ordered phase (Fig.\ref{fig:FM-AFOP}(b)). \\
%
\begin{figure}
\includegraphics[width=0.99\columnwidth]{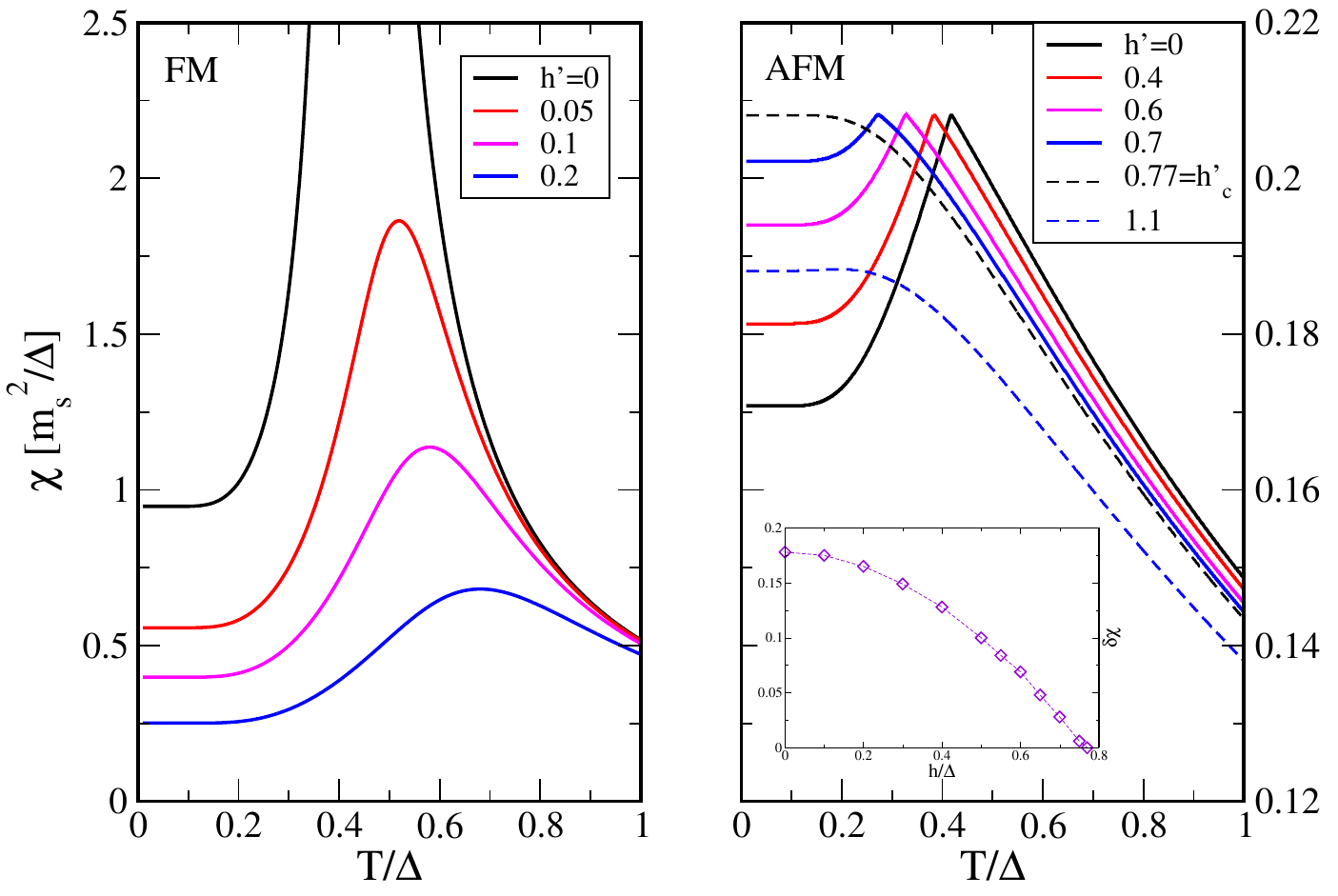}
\caption{Temperature variation of FM and AFM {\it homogeneous} susceptibilities for different fields for $\xi_s=1.2$. left panel: FM zero-field divergence at $T_m/\De$=0.42 is rapidly suppressed by external field. right panel: AF anomaly at $T_m(h')$ (see also Fig.~\ref{fig:AF-Tm}(a)) persists but shifts with lower critical temperatures and vanishes at critical field $h'_c=h_c/\De$=0.77. Inset shows relative susceptibility drop (Eq.~(\ref{eq:dropsus})) with increasing field, starting at $\delta_{AFM}(0)$=0.18. Dashed lines correspond to critical and larger fields.}
\label{fig:FMAF-susz-t}
\end{figure}
%
Without external field ($h'$= $0$) the order parameter $\la S_x\ra_\lam=\la S_x\ra$ is sublattice independent
for FM and opposite  $\la S_x\ra_\lam=\lam\la S_x\ra$ in AFM case. Then in the zero-field 
case Eq.~(\ref{eq:MF})  reduces to
\bea
\hDe_T=\xi_sf_s(T);\;\; f_s(T)=\tanh\bigl[(\frac{\De}{2T})\hDe_T\bigr],
\label{eq:aux}
\eea
now with simplified $\hDe_T=[1+(2\xi_s\la S_x\ra_T)^2]^\fs$ and $\la S_x\ra_T=\frac{1}{2\xi_s}[\xi_s^2f_s^2(T)-1]^\fs$ leading to a saturated splitting $\hDe_0=\xi_s$ and moment  $\la S_x\ra_0=\frac{1}{2\xi_s}(\xi_s^2-1)^\fs$ that increases with singular slope above the QCP $\xi^c_s=1$ (see Ref.\onlinecite{thalmeier:24}).
In both cases at the transition temperature $\hDe_{T_m}=1$ or $f_s(T_m)=\xi_s^{-1}$ which leads to the
explicit zero-field solution \cite{thalmeier:24}:
\bea
T_m(\xi_s)&=&\frac{\De}{2\tanh^{-1}\bigl(\frac{1}{\xi_s}\bigr)}.
\label{eq:tm0}
\eea
This means that an induced magnetic order with finite $T_m$ at zero field (for FM as well as AFM) appears only for $\xi_s>\xi_c=1$ which  marks the QCP between paramagnetic $(\xi_s<\xi_c)$ and magnetic $(\xi_s>\xi_c)$ ground states. For finite field there is no transition temperature in the FM case but only a crossover behaviour. In the AFM case the phase transition prevails and $T_m(h)$, i.e., the $\text{h-T}$ phase diagram can be obtained from the numerical solution of Eq.~(\ref{eq:MF}). For  FM at all $T$ and for AFM for  $T>T_m(h')$ $(M_x^s=0)$ the homogeneous magnetisation $M_x^0(h')$ is also obtained from this equation simply by replacing
$\la S_x\ra_\lam,\la S_x\ra_{\blam}\rightarrow M_x^0$ with $\sigma=\pm$ for FM and AFM , respectively.
%
\begin{figure}
\includegraphics[width=0.95\columnwidth]{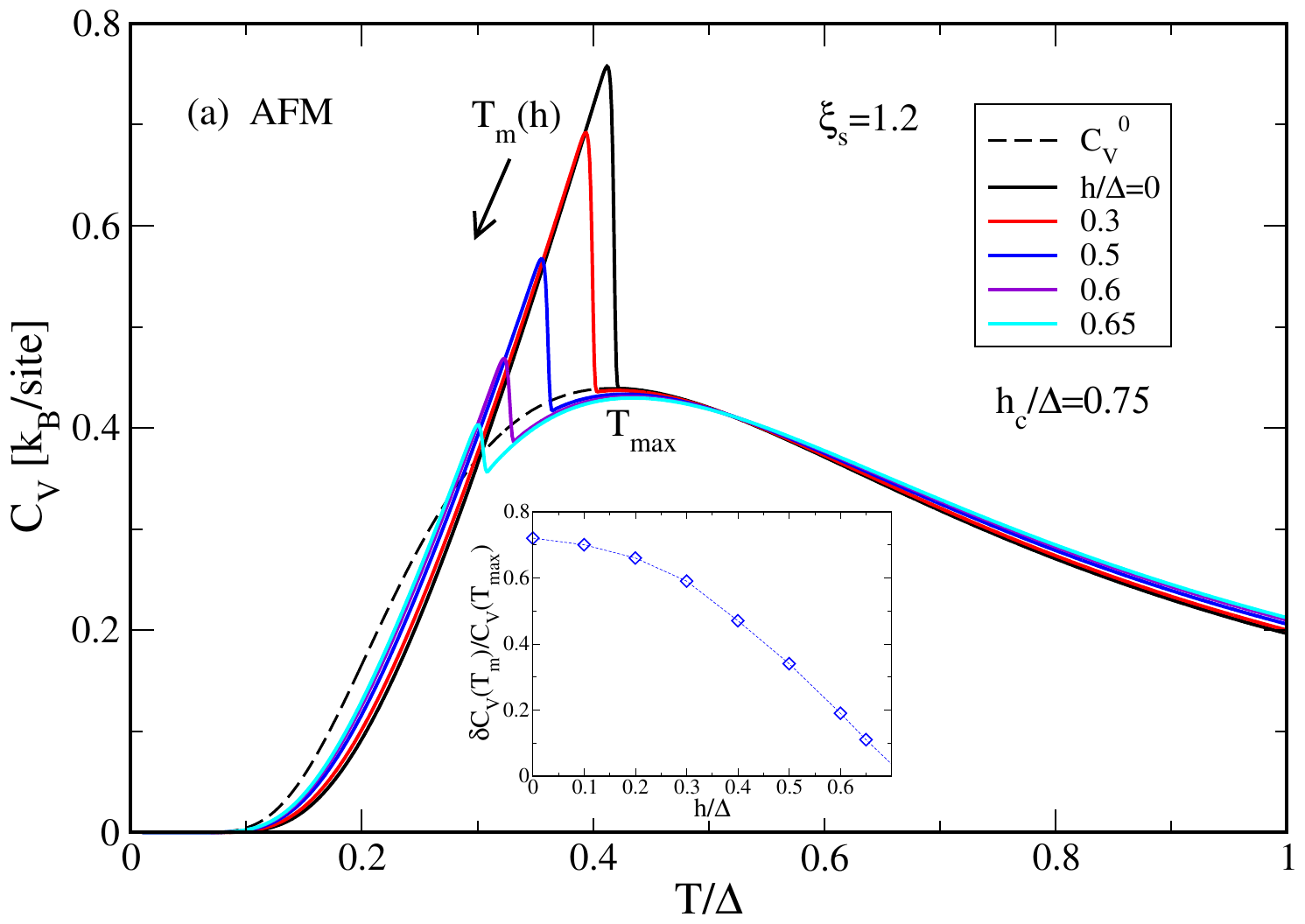}
\includegraphics[width=0.95\columnwidth]{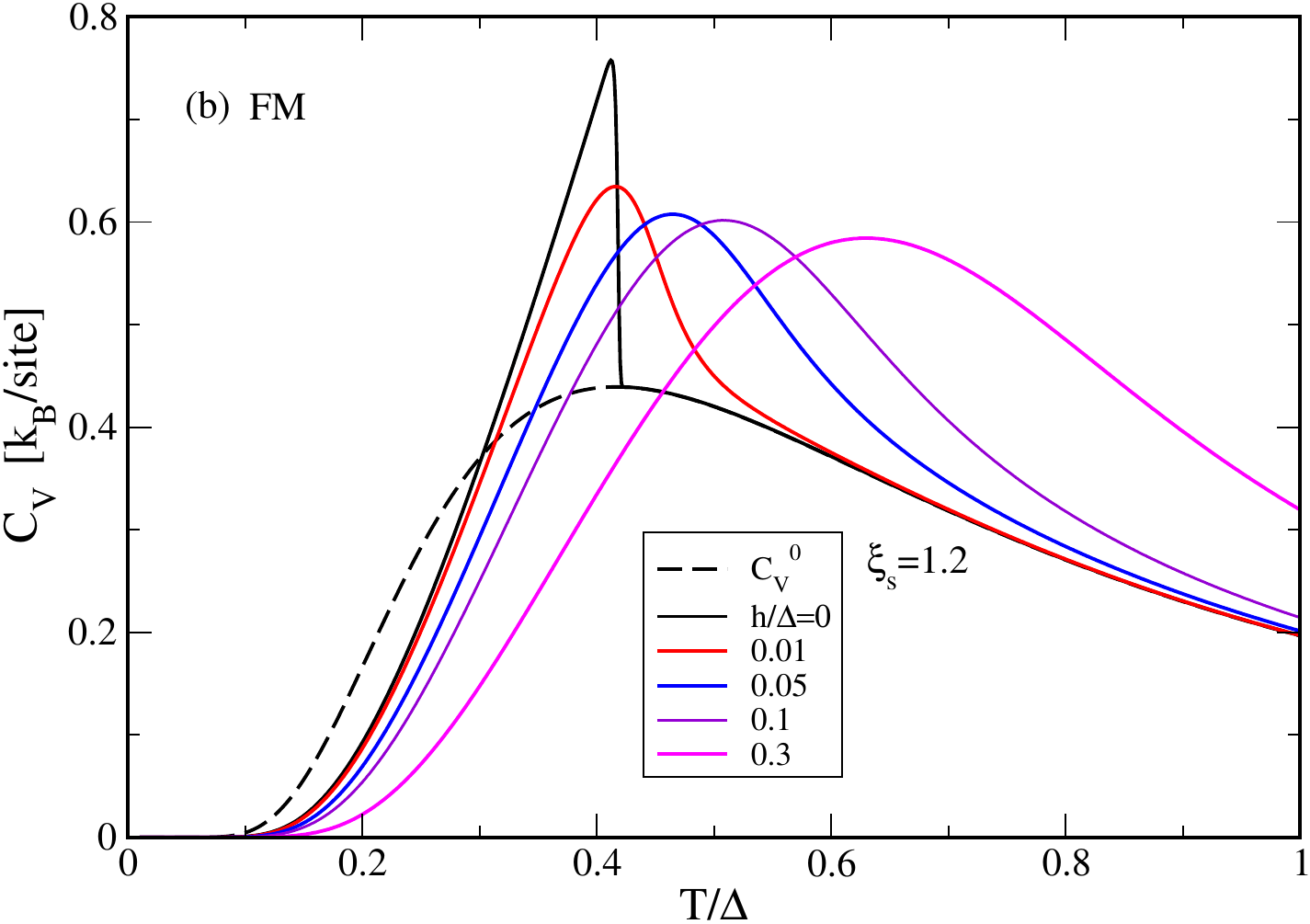}
\caption{(a) Specific heat for AFM case with $\xi_s=1.2$ where ordering temperature $T_m(h=0)$ coincides with
maximum temperature $T_{max}=0.42\De$ of the Schottky background anomaly. For increasing 
field $T_m(h)$ decreases (Fig.~\ref{fig:AF-Tm}(a)) and concomitantly the relative step size $\delta C_V(T_m(h))/C_V(T_{max})$ decreases (inset); they both vanish at the critical field $h_c(\xi_s)$.
(b) Specific heat for FM with $\xi_s=1.2$ which is identical to AFM  for $h=0$. However 
for finite field the FM order changes rapidly into a crossover (cf. Fig.~\ref{fig:FM-AFOP}(a)) and the sharp jump broadens
and merges with the Schottky background for larger field. This type of field dependence should be expected for
the induced FM \PR~shown in Fig.~\ref{fig:gacomp} (only for zero field case).
.}
\label{fig:AFFM-cv-t}
\end{figure}
%
The order parameters, uniform and staggered magnetisations for the two cases and AFM transition temperature   $T_m(h,\xi_s)$ and the critical field $h_c(\xi_s)$ are  shown in the Figs.\ref{fig:FM-AFOP},\ref{fig:AF-Tm} which will be discussed in Sec.~\ref{sec:discussion}.

\subsection{Static field dependent susceptibility}

An alternative approach to obtain $T_m(h)$  and the critical field $h_c(\xi_s)$ defined by $T_m(h_c)$=$0$ proceeds
via the determination of collective  static RPA susceptibilities.
In the Ising-type singlet-singlet system only the operator $J_z=m_sS_x$ has matrix elements and leads
to a finite static single-site susceptibility 
. For finite field and temperature it is given by \cite{jensen:91,thalmeier:24} (sublattice index $\lam$ suppressed, $\tau,\tau'=\pm$):
\bea
\chi_0(T,h')&=&m_s^2\sum_{\tau\neq\tau'}|\la \tau|S_x|\tau'\ra |^2\frac{p_\tau-p_{\tau'}}{\eps_{\tau'}-\eps_\tau}\non\\
&+&\beta\bigl[\sum_\tau |\la \tau|S_x|\tau\ra |^2-\la S_x\ra|^2\bigr],
\label{eq:statsus0}
\eea
consisting of a first vanVleck term and and a second pseudo-Curie term induced by the order parameter. This may be evaluated using the pseudo spin matrix in the basis of eigenstates $|\tau\ra$ given in Eq.(\ref{eq:eigenstate}) as
\bea
\{S_x\}_{\tau\tau'}^\lam=
\frac{1}{2}\left(
 \begin{array}{cc}
\sin 2\theta_\lam& \cos 2\theta_\lam\\
 \cos 2\theta_\lam&-\sin 2\theta_\lam\\
\end{array}
\right).
\label{eq:sxmat}
\eea
Then the total static susceptibility, expressed by the sum of vanVleck and pseudo Curie terms  $\chi_{0\lam}=\chi_{0\lam}^{vV}+\chi_{0\lam}^C$ which are explicitly obtained as
\bea
\chi_{0\lam}^{vV}(T,h')&=&\frac{m_s^2}{2\De}\frac{1}{(\hDe_T^\lam)^3}\tanh\bigl[(\frac{\De}{2T})\hDe^\lam_T\bigr]\non\\
\chi_{0\lam}^{C}(T,h')&=&(\fs m_s)^2\frac{1}{T}(1-\frac{1}{(\hDe_T^\lam)^2})(1-f_s^{\lam 2}).
\label{eq:sus0}
\eea
In the general FM and the paramagnetic AFM case  there is no sublattice ($\lam$=A,B) dependence.
Note that the pseudo Curie term vanishes for $T\rightarrow 0$ despite the $1/T$ Curie prefactor because the subsequent population factors suppress it exponentially. From the (generally sublattice dependent) single ion susceptibility  $\chi_{0\lam}$ we obtain the momentum $\bq$ dependent RPA susceptibility of the interacting
system. The Fourier transform of the n.n. exchange interaction in Eq.~(\ref{eq:Ham}) is given by $I(\bq)=zI_0\gamma(\bq)$ with $\gamma(\bq)=z^{-1}\sum_{\la ij\ra} \exp(\bR_i-\bR_j)$ denoting the structure function for the (Bravais) lattice. Then the collective static susceptibility for the FM is given by
%
\begin{figure}
\includegraphics[width=0.95\columnwidth]{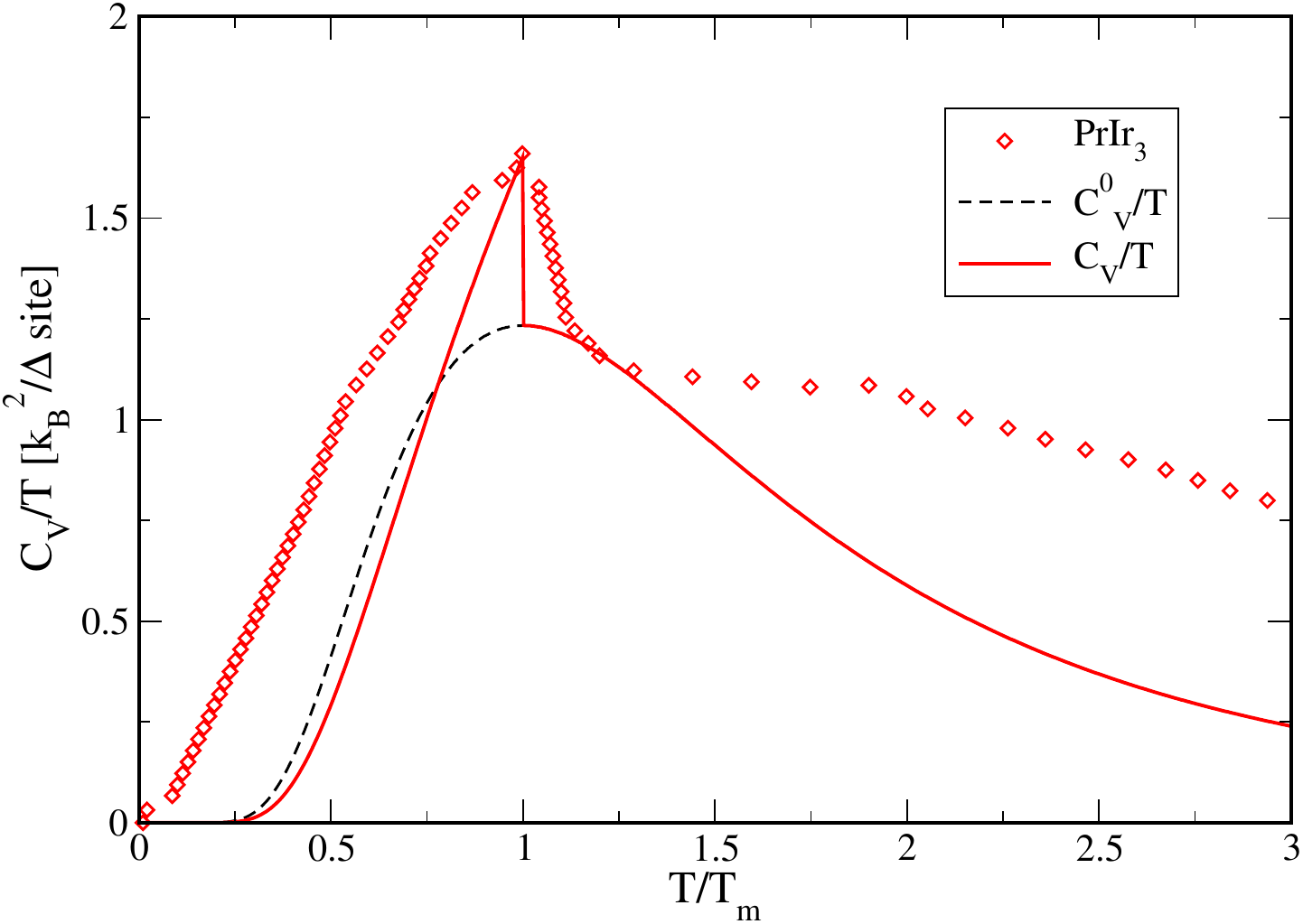}
\caption{Specific heat coefficient $\gamma = C_V/T$ of the the close to critical singlet-singlet induced ferromagnet at zero field. Dashed black line is the background Schottky anomaly and red line corresponds to control parameter $\xi_s=1.08$ close to the quantum critical $\xi^c_s$= 1.
The jump at FM $T_m\equiv T_C$ appears at the maximum postion $T_{max}$ of the Schottky background $C^0_V/T$ (dashed black line). For comparison  the magnetic specific heat of \PR~(T$_C$=7.5 K) adapted from Ref.~\onlinecite{gornicka:24} is shown by the red symbols (with residual T=0 $\gamma$-value  subtracted and the maximum position and value scaled to coincide).}
\label{fig:gacomp}
\end{figure}
%
%
\bea
\text{FM}:\;\;&&\chi(\bq)=\chi_{0}[1-I(\bq)\chi_{0}]^{-1},
\label{eq:statsuFM}
\eea
and the susceptibility matrix for the AFM (in sublattice A,B space) is derived as
\bea
\hspace{-0.4cm}
\text{AFM}:
\hat{\chi}(\bq)&=&
\frac{1}{D(\bq)}\left(
 \begin{array}{cc}
\chi_{0A}& I(\bq)\chi_{0A}\chi_{0B}\\
 I(\bq)\chi_{0B}\chi_{0A}&\chi_{0B}\\
\end{array}
\right)\non\\[0.2cm]
D(\bq)&=&1-I(\bq)^2\chi_{0A}\chi_{0B}.
\label{eq:statsuAF}
\eea
%
The physical homogeneous susceptibility for the AFM is then obtained as the average over sublattice components according to 
\bea
\chi(0)=\fs\sum_\lam\chi_{0\lam}[1+I_e\chi_{0\lam}]^{-1}.
\label{eq:homsuAF}
\eea
As zero-field  limiting cases we obtain for T= 0, T$_m$ the following values \cite{thalmeier:24} (in units of $[m_s^2/\De$]):
For FM $\chi(0)=1/[2\xi_s(\xi_s^2-1)]$ and $\chi(T_m)\rightarrow\infty$ while for AFM $\chi(0)=1/[2\xi_s(\xi_s^2+1)]$ and
$\chi(T_m)=1/(4\xi_s)$. Note this is the 'longitudinal' zz- type susceptibility shown in Fig.~\ref{fig:FMAF-susz-t} (there is no transversal one in this Ising-type model) and it stays finite for T$\rightarrow 0$. This is different from degenerate ground state magnets where it vanishes in that limit. For AFM the relative drop from the maximum at T$_m$ may be expressed as
\bea
\delta\chi_{AFM}(h)&=&\frac{\chi(T_m,h)-\chi(0,h)}{\chi(T_m,h)}\non\\
\delta\chi_{AFM}(0)&=&(\xi_s^2-1)/(\xi_s^2+1).
\label{eq:dropsus}
\eea
This experimentally accessible quantity provides a direct measure for the control parameter in the AFM two-singlet induced magnet. Another one applicable to both cases will be discussed below in connection with the specific heat anomalies. In the degenerate limit when $\xi_s\gg 1$ the conventional complete drop with $\delta\chi_{AFM}(0)\approx 1$ is achieved.\\

Starting from the  paramagnetic  side of the AFM without sublattices  we also may determine the  transition temperature $T_m(h')$ at finite field or AFM phase boundary from the condition that the AFM staggered  susceptibility (\bq=\bQ) diverges,
i.e. $I(\bQ)\chi_0(T_m)=1$ with $I(\bQ)=z|I_0|=I_e$ where $\bQ$ is the AFM ordering vector. Using Eq.~(\ref{eq:sus0})
this leads to the implicit solution for $T_m(h')$ according to
\bea
f_s(\hT_m)=\frac{\hDe_{T_m}^3}{\xi_s}-\fs\frac{\hDe_{T_m}}{\hT_m}(\hDe_{T_m}^2-1)\bigl[1-f_s^2(\hT_m)\bigr],
\label{eq:tmh}
\eea
with the normalized $\hT_m=T_m(h')/\De$ and  paramagnetic level splitting $\hDe_{T_m}(h')=[1+(m_sh'+2\xi_s\la S_x\ra)^2]^\fs$. The first and second terms on the r.h.s. of this equation result from vanVleck and pseudo Curie terms of the susceptibility, respectively. In zero field limit the latter vanishes and $\hDe_{T_m}$=1, then the explicit solution of Eq.~(\ref{eq:tm0}) is recovered.

Furthermore the critical field for AFM order for a given control parameter 
$\xi_s > 1$ {\BLU  is defined by the condition $T_m(h'_c)$= 0. At zero temperature the
population difference in Eq.(\ref{eq:auxlam}) reaches its maximum, meaning  $f_s(\hT_m(h'_c))$= 1.
Inserting this into Eq.~(\ref{eq:tmh}) leads to the implicit equation for the critical field according to}
\bea
h_c'(\xi_s)=\frac{1}{m_s}\bigl[(\xi_s^\frac{2}{3}-1)^\fs+2\xi_s\la S_x\ra_0\bigr],
\eea
which starts with a singular slope at the QCP $\xi_s=1$.
Indeed close to the QCP with $\xi_s=1+\delta$ $(\delta\ll 1)$ a comparison of the asymptotic zero-field $T_m(\delta)$ and $h'_c(\delta)$ leads to the result
\bea
T_m(\delta)/\De&\simeq&|\ln\frac{\de}{2}|\non\\
h'_c=h_c(\de)/\De&\simeq&\frac{\sqrt{2}}{m_s}(1+\frac{1}{\sqrt{3}})\sqrt{\de}.
\label{eq:critasymp}
\eea
The logarithmic and square root singularities of these critical quantities can be seen in the comparison with the numerical
calculations presented in Fig.~\ref{fig:AF-Tm}(b).
%
\begin{figure}
\vspace{-0.2cm}
\includegraphics[width=1.04\columnwidth]{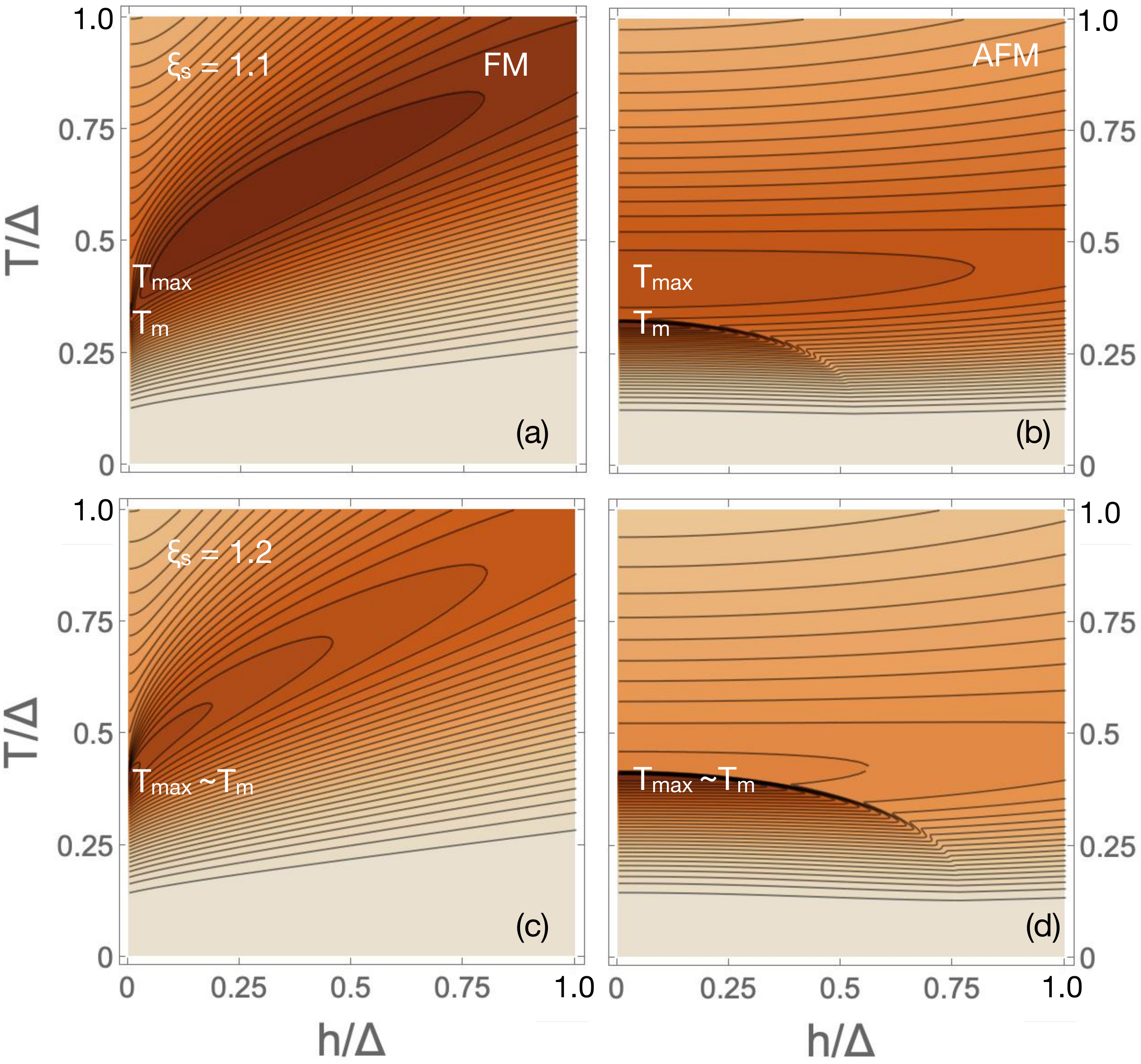}
\caption{Contour plots of FM (a,c) and AFM (b,d) specific heat $C_V(T,h)$ for two different control parameters $\xi_s$
with zero-field transition temperatures $T_m(\xi_s=1.1)=0.33$ and  $T_m(\xi_s=1.2)=0.42$.
In the FM case a transition occurs only for $h$= 0 and changes into a crossover with the Schottky type specific heat maximum at $T_{max}(h)$
shifting to higher temperature. In the AFM case the transition persists at all fields at $T_m(h)$ up to the critical fields
$h'_c(\xi_s=1.1)=0.53$  and  $h'_c(\xi_s=1.2)=0.77$ (cf. Fig.~\ref{fig:AF-Tm}). The relative zero-field position of $T_m$ and background $T_{max}$ 
depends on $\xi_s$ (it coincides for $\xi_s=1.2$ in (d)) and they separate on approaching the critical field.
Equidistant contour lines
of $C_V(T,h)$ vary between 0-0.5 k$_B$/site for $x_s=1.1$ and 0-0.75 k$_B$/site for $\xi_s=1.2$ 
(see Fig.~\ref{fig:AFFM-cv-t}).}
\label{fig:CVcont}
\end{figure}
%
\section{Caloric properties and coefficients}
\label{sec:caloric}

With the basic properties of the model derived above we can now calculate the thermodynamic potentials
as a prerequisite for obtaining the various thermodynamic quantities relevant for experimental investigation of 
singlet induced moment magnets. These potentials depend on the set of variables $(T,h,\vare)$ where T is the temperature, h the external field and $\vare$ is the volume strain $\vare$=$\frac{1}{3}$($\vare_{xx}$+$\vare_{yy}$+$\vare_{zz}$)  which enters through the magnetoelastic strain dependence $\Delta(\vare)$ of the singlet-singlet splitting.\\

\subsection{The thermodynamic potentials}
\label{sec:thermpot}

Firstly we calculate the internal energy which has two contributions according to Eq.~(\ref{eq:MFHam}),
coming from the thermal averaging over level energies $(U_0)$ and a direct order parameter term $(U_{op})$,
respectively. The internal energy per site is then given by
\bea
&&U(T,h')=U_0+U_{op}\non\\
&&=\fs\sum_\lam\sum_n p^n_\lam\eps^n_\lam
+\fs\sigma m_s^2 I_e \la S_x\ra_\lam \la S_x\ra_{\blam},
\eea
where the thermal occupations of MF energy levels are $p^n_\lam=Z_\lam^{-1}\exp(-\beta\eps^n_\lam)$
with $Z_\lam=2\cosh\bigl[(\frac{\De}{2T})\hDe_T^\lam\bigr]$ denoting the partition function in each sublattice; this leads to 
\bea
U(T,h')=&&-\frac{\De}{4}
\sum_\lam\hDe^\lam_T\tanh\bigl[(\frac{\De}{2T})\hDe^\lam_T\bigr]\non\\
&&+\fs\sigma m_s^2 I_e \la S_x\ra_\lam \la S_x\ra_{\blam}.
\label{eq:uint}
\eea
%
\begin{figure}
\includegraphics[width=0.95\columnwidth]{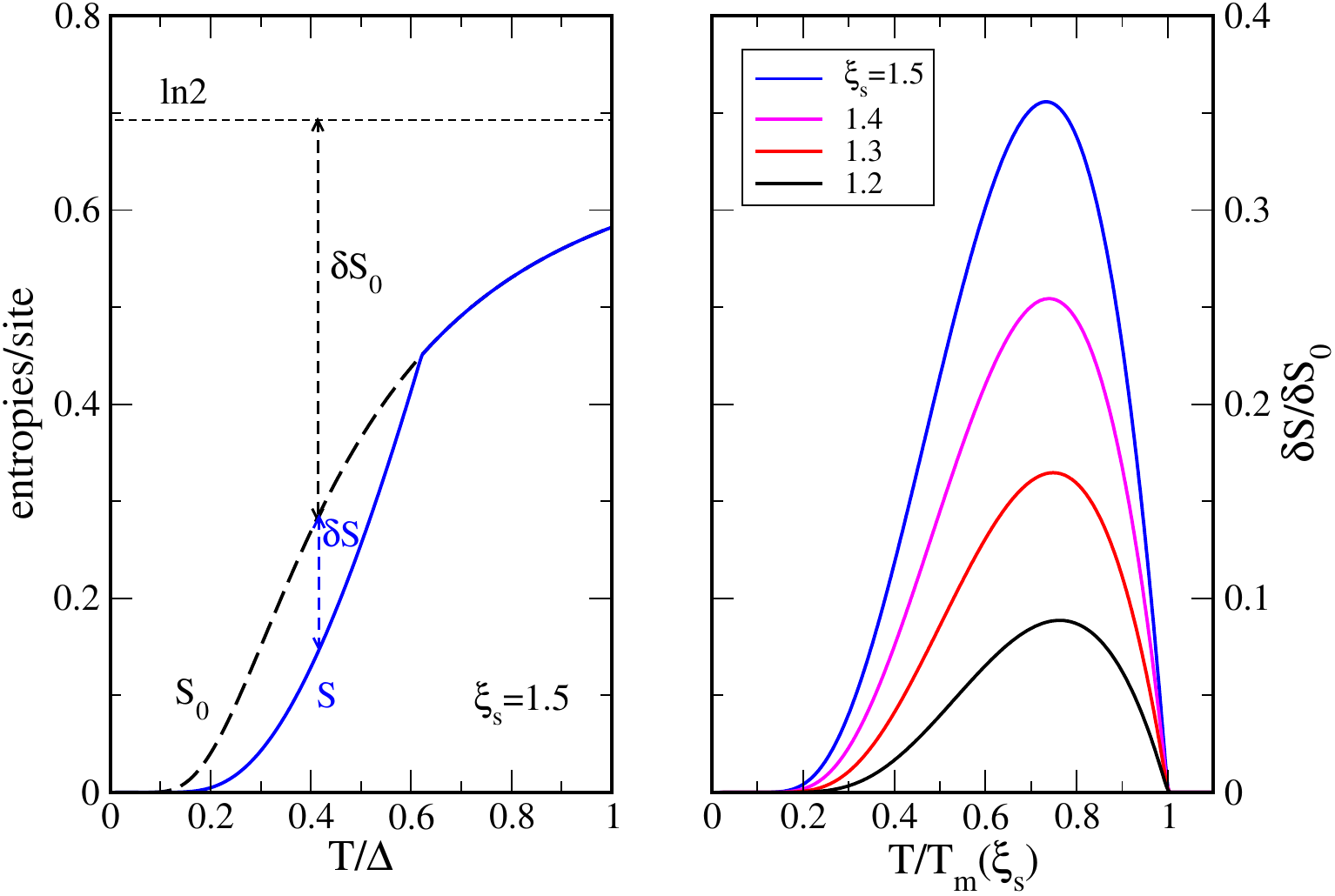}
\caption{left panel: Two contributions to the entropy release: $\delta S_0=\ln2-S_0$ from singlet depopulation and
$\delta S=S_0-S$ from induced magnetic order.
right panel: Relative entropy release $\delta S/\delta S_0$ at {\it zero field} from the  transition (FM or AFM)  compared to the one from excited singlet depopulation. It  increases with the control parameter $\xi_s$ which shifts the transition $T_m(\xi_s)$ to higher temperatures. For $\xi_s\gg 1$ the transition exhausts all the entropy release $\ln 2$. }
\label{fig:relent}
\end{figure}
%
Using the results of the previous section, in particular Eq.~(\ref{eq:MF}) $U(T)$ may be written more explicitly for
the FM case as
\bea
\hspace{-0.3cm}
U(T,h')&=&-\frac{\De}{2}\bigl\{\hDe_Tf_s(T)-\frac{1}{2\xi_s}[(\hDe_T^2-1)^\fs-m_sh']^2\bigr\}\non\\
\hDe_T&=&[1=(m_sh'+2\xi_s\la S_x\ra)^2\non\\
\la S_x\ra&=&\frac{\hDe_T}{2}(m_sh'+2\xi_s\la S_x\ra)\tanh\bigl[(\frac{\De}{2T})\hDe_T\bigr].
\label{eq:unitFM}
\eea
In the AF case it is preferable to use the general two-sublattice expression in Eq.~(\ref{eq:uint}) together with Eq.~(\ref{eq:MF}). 
In the zero-field limit  the internal energy for FM and AFM reduces to the single expression
\bea
U(T,0)=-\frac{\De}{4}\bigl[\xi_sf_s^2(T)+\frac{1}{\xi_s}\bigr],
\label{eq:internal}
\eea
leading to a ground state energy $U_{gs}=-\frac{\De}{4}(\xi_s+\frac{1}{\xi_s})$.\\

Furthermore the entropy of the model is given by
\bea
\hspace{-0.4cm}
S=\frac{U_0}{T}+ln Z
=-\bigl(\frac{\De}{2T}\bigr)\sum_\lam\hDe^\lam_Tf_s^\lam(T)
+\fs\sum_\lam Z_\lam,
\label{eq:entropy0}
\eea
which may be evaluated explicitly as
\bea
S(T,h')&=&\fs\sum_\lam\Bigl\{\ln\Bigl[2\cosh\bigl[\bigl(\frac{\De}{2T}\bigr)\hDe_T^\lam\bigr]\Bigr]\non\\
&&-\bigl(\frac{\De}{2T}\bigr)\hDe^\lam_T\tanh\bigr[\bigl(\frac{\De}{2T}\bigr)\hDe_T^\lam\bigr]\Bigl\},
\label{eq:entropy1}
\eea
and approaches the limit $S=\ln 2$ for $T\gg\De$.\\

Finally for the free energy per site  $F=U-TS=\\
U_{op}-T\fs\sum_\lam Z_\lam$ we obtain explicitly:
\bea
F(T,h')=&&
-T\fs\sum_\lam\ln\Bigl[2\cosh\bigl(\frac{\De}{2T}\bigr)\hDe_T^\lam\Bigr]\non\\
&&+\fs\sigma m_s^2 I_e \la S_x\ra_\lam \la S_x\ra_{\blam}.
\label{eq:free}
\eea
The free energy is needed to calculate the thermal expansion coefficient that enters 
the expression for the barocaloric cooling rate.
%
\begin{figure}
\includegraphics[width=0.80\columnwidth]{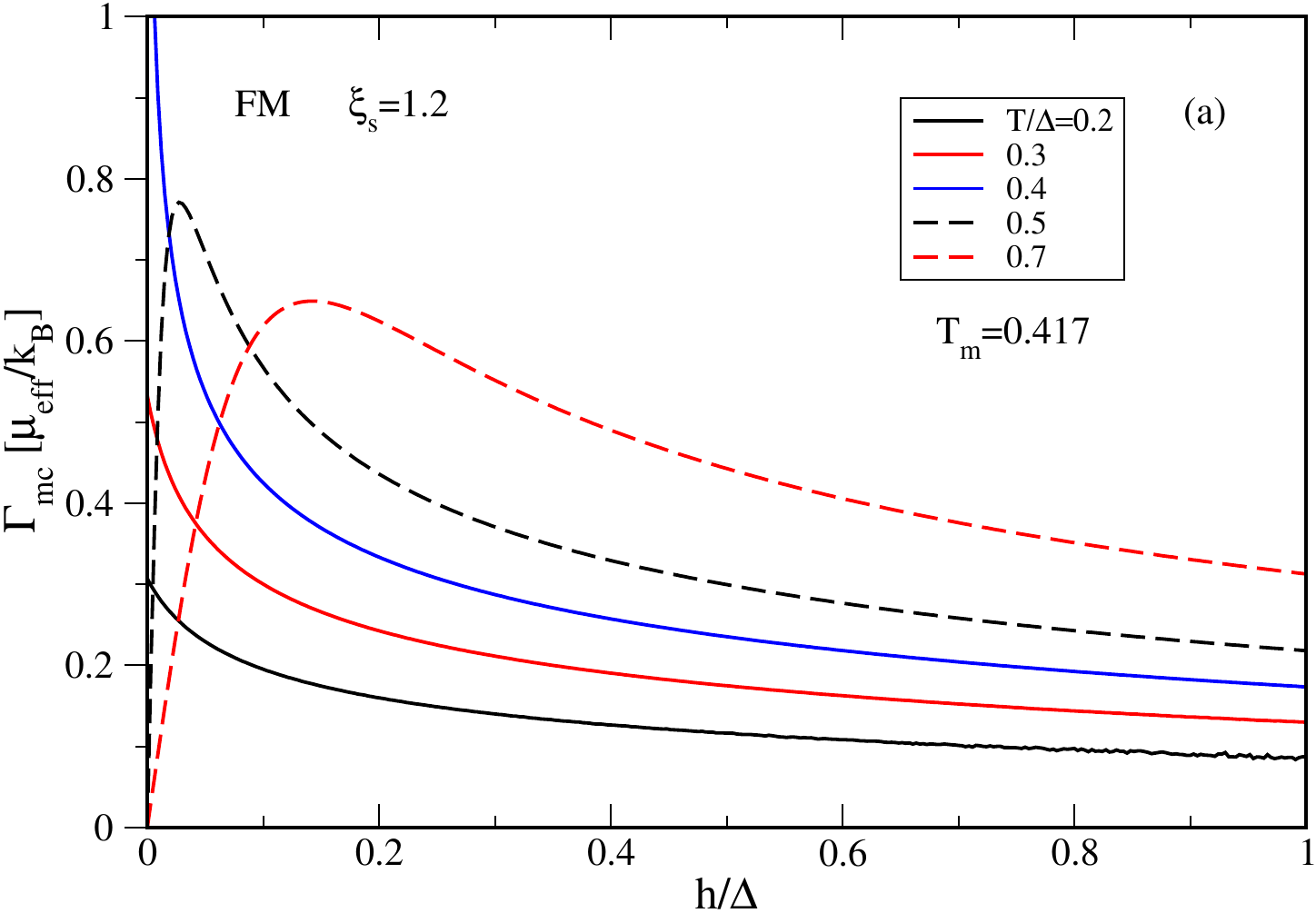}\\[0.2cm]
\includegraphics[width=0.80\columnwidth]{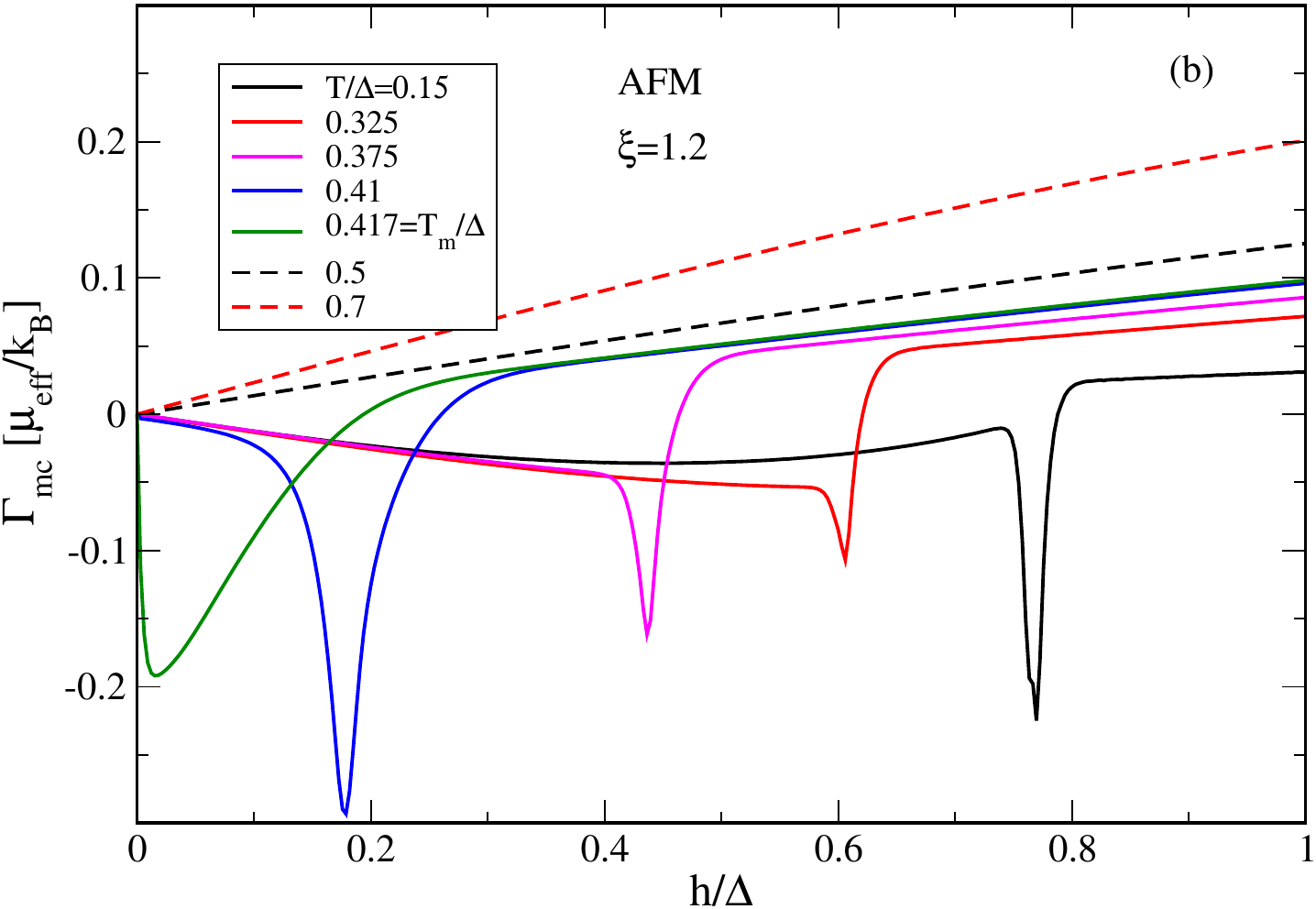}
\caption{Magnetocaloric coefficients (adiabatic cooling rate with magnetic field) for FM and AFM cases with $\xi_s=1.2$. (a) For FM a monotonic increase for lowering the field is observed $(T<T_m)$ which diverges on reaching $T_m$ from below. For temperatures $T>T_m$ (dashed) it will be suppressed to zero in the low field limit.
{\BLU (b) For AFM  a sign change and downward peak occurs at the 
field where T$_m$(h)=T, i.e., when the phase boundary (Fig.~\ref{fig:AF-Tm}(a)) is crossed (T given in the legend). 
At the lowest fields $\Gamma_{mc}$  approaches zero. For temperatures above the value T$_m$(h=0)$/\De=0.417$  (dashed) a featureless sub- linear field dependence occurs.}
}
\label{fig:mce}
\end{figure}
%

\subsection{Specific heat, magneto- and barocaloric coefficients and thermal expansion}
\label{sec:calcoeff}

The field- and temperature dependence of the specific heat under constant volume (strain) is obtained
from the internal energy via the general thermodynamic relation
\bea 
C_V(T,h')=\bigl(\frac{\partial U(T,h')}{\partial T}\bigr)_\vare,
\eea
\\
where the relation between volume $V$ and volume strain $\eps$ is given by $\vare=(V-V_0)/V_0$ with $V_0$ denoting the
ambient pressure equilibrium value .
The temperature and field dependence  has to be computed numerically from the internal energy in Eq.~(\ref{eq:uint})
after solving the selfconsistency equations for the order parameter (Eq.~(\ref{eq:MF})). In the zero field case one 
can make qualitative conclusions: The entropy release as discussed in Sec.~\ref{sec:discussion} can happen by i) the depopulation of the upper singlet leading to the underlying Schottky specific heat peak and ii)
by the induced order leading to the specific heat jump at T$_m(\xi)$ {\BLU (Fig.\ref{fig:AFFM-cv-t})}. The former is given by $C_V(T)=x^2/ \cosh^{2}x$ [k$_B$/site] with $x=\De/2T$. Its maximum position, determined by $x\tanh x$ =$1$, lies at $T_{max}=0.42\De$ (with C$_V$(T$_{max}$) = 0.44k$_B$) while
the (zero-field) transition temperature $T_m$ is given by Eq.~(\ref{eq:tm0}). Their observed relative position then allows to draw conclusions on the size of the control parameter $\xi_s$ (Sec.~\ref{sec:PrIr3}), see also Fig.~\ref{fig:AF-Tm}.

Further interesting and important experimentally accessible quantities are the magneto- and elasto- (or baro-) caloric coefficients, i.e. the adiabatic (S=const) cooling rates under rapid field or hydrostatic pressure change, respectively
\cite{straessle:03,gati:23,zic:25}. {\BLU These effects have also technical applications which are not in the focus in this work and we refer to Refs.\onlinecite{pecharsky:99,sandeman:12} for a deeper discussion of this important issue.}
Microscopically the magneto- and barocaloric coefficient are due to the field dependence of non-diagonal elements in the Hamiltonian of Eq.~(\ref{eq:MFHam}) or the volume strain dependence of diagonal terms (i.e. the CEF splitting), respectively. The adiabatic magnetocaloric  coefficient is given by
\bea
\Gamma_{mc}(T,H)=\bigl(\frac{\partial T}{\partial H}\bigr)_S=
-\frac{T}{C_V}\bigl(\frac{\partial M_x}{\partial T}\bigr)_H,
\label{eq:mce}
\eea
with the homogeneous magnetisation $M_x$ = $m_s(g_J\mu_B)M_x^0$ = $(g_{eff}\mu_B)M_x^0$ and $M_x^0$ defined in Eq.~(\ref{eq:mag}).  Likewise we have for the adiabatic barocaloric coefficient at a constant field H
\bea
\Gamma_{bc}(T,H)=\bigl(\frac{\partial T}{\partial p}\bigr)_S=
\frac{T}{C_{p}}\bigl(\frac{\partial \vare}{\partial T}\bigr)_p,
\label{eq:bce}
\eea
where $p=-c_B\vare$ is the pressure and the last factor in the above equation is the thermal expansion coefficient $\beta$
with  $c_B$ denoting the inverse bulk modulus or fully symmetric elastic constant.
\bea
\beta(T,H)=\bigl(\frac{\partial\vare}{\partial T}\bigr)_p;\;\;
\epsilon(T)=-\frac{1}{c_B}\bigl(\frac{\partial F}{\partial\vare}\bigr)_T.
\label{eq:striction}
\eea
These are also an important accessible quantities in CEF systems \cite{novikov:17} and can be obtained from the free energy as indicated. We assume  that the strain coupling appears predominantly through the volume strain dependence of the CEF splitting
$\De(\epsilon)$. Then a dimensionless CEF Gr\"uneisen parameter $\Omega=-(\partial ln \De(\epsilon)/\partial \epsilon)$  (which may be positive or negative) characterises the strength of this coupling. In rare earth compounds this parameter is usually of the order of $|\Omega|\approx$ 10. \cite{thalmeier:91,novikov:17}. Then we may write for the strain dependent energy scale $\De(\epsilon)=\De(1-\Omega\epsilon)$ and control parameter $\xi_s(\epsilon)=\xi_s(1+\Omega\epsilon)$ and this leads to the strain dependence of the free energy in Eq.~(\ref{eq:free}) which determines thermal expansion, magnetostriction and barocaloric coefficients given in the equations above. 

Typical examples of the magneto- and barocaloric quantities for FM and AFM case are discussed in Sec.~\ref{sec:discussion}.

%
\begin{figure}
\includegraphics[width=0.95\columnwidth]{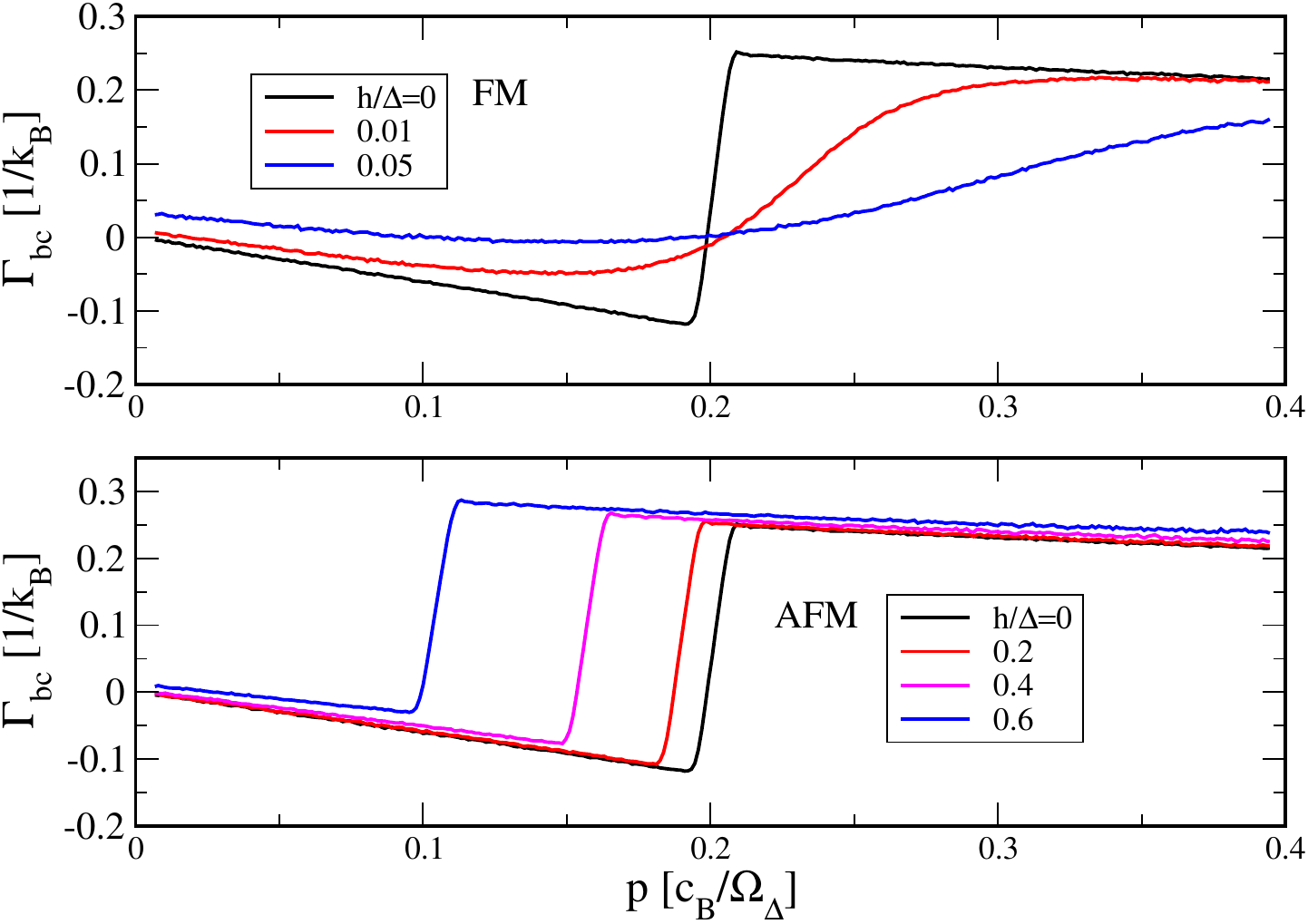}
\caption{Barocaloric coefficient (adiabatic cooling rate with pressure) for FM and AFM cases with $\xi_s=1.3$ ( $T_m/\De=0.491$) and $T/\De=0.3$. For zero field both are identical with sign changing step at critical pressure $p_c\sim 0.2$.
For constant finite field $\Gamma_{bc}$ becomes smooth while for AFM case the critical pressure moves to lower values
with increasing field but the pressure dependence keeps its step like shape.}
\label{fig:bce}
\end{figure}
%
%
%

{\BLU
\section{Material examples for singlet induced moment magnets}
\label{sec:materials}
}

{\BLU
Now we give a brief overview of materials where the theory of field dependent properties in  singlet ground state induced quantum magnets developed here may be relevant and then apply it to a most recent example of the near critical ferromagnet \PR. 

In this overview we also include materials where the first excited CEF state is not a singlet but may also be a doublet (for four- and sixfold uniaxial point groups) or a triplet (for tetrahedral and cubic point groups only). It was demonstrated in Ref.~\onlinecite{thalmeier:24} that their principal thermodyamic properties are similar to the singlet-singlet case. As shown in the following section the induced moment magnetism may also be preceded in the paramagnetic phase by a strong temperature dependence and a softening of a critical magnetic exciton mode to a varying degree at the ordering wave vector. A considerable number of materials have been accumulated that qualify as singlet ground state  induced moment system. Frequently these compounds contain non-Kramers 4f or 5f ions like (J=4)  Pr$_3$Tl\cite{buyers:75,birgeneau:71},  PrSb~\cite{mcwhan:79}, PrCu$_2$\cite{kawarazaki:95}, PrNi\cite{savchenkov:19}, Pr$_5$Ge$_4$ \cite{rao:04},  Pr$_2$CuO$_4$ \cite{sumarlin:95}  and elemental  Pr- metal under pressure~\cite{jensen:91,birgeneau:72,cooper:72,jensen:87}. Singlet ground state induced AFM order has also been found in Tb and Tm compounds (both with J=6) TbSb~\cite{holden:74} and its Y- substituted alloy \cite{cooper:70} as well as Tb$_3$Ga$_5$O$_{12}$ \cite{wawr:19} and KTmSe$_3$ \cite{zheng:23}. Recently a 4f singlet-singlet system PrMgAl$_{11}$O$_{19}$ \cite{kumar:25} has been discovered with Pr located on a triangular sublattice. This raises the interesting theoretical question how the induced moment mechanism interferes with geometric frustration. Indeed this compound does not order magnetically down to $0.4 $ K.

Furthermore candidates with 5f electrons  for induced magnetism are UGa$_2$~\cite{marino:23a} and UPd$_2$Al$_3$~\cite{grauel:92,mason:97,thalmeier:02} and URu$_2$Si$_2$\cite{sundermann:16} (further references cited therein), and also Fe-substituted~\cite{marino:23b}. However, it should be noted that the degree of itineracy of 5f states is often higher and the localised picture may not be fully applicable. When considered within the localised 5f scenario URu$_2$Si$_2$\cite{sundermann:16,marino:23b} and also the (Kramers ion) compound YbRu$_2$Ge$_2$ \cite{jeevan:06} show that the induced order mechanism is not restricted to magnetism but also possible for multipolar~\cite{santini:94,haule:10} and (ferro-)quadrupolar~\cite{takimoto:08,rosenberg:19} order, respectively.

Finally another aspect of f- electron materials with CEF singlet ground state has been discovered: For suitable 2D lattice structure with a basis like honeycomb or kagome the magnetic excitons may develop a nontrivial topological character with non-vanishing Chern number which requires the presence of magnetic excitonic edge modes in the paramagnetic state~\cite{akbari:23}. An example is the 2D honeycomb lattice with a two-atom basis with site symmetry $C_{3v}$. which may be realised as a planar structure within a 3D lattice.
This structure is relevant for various f-electron compounds like Na$_2$PrO$_3$ \cite{daum:21}, TmNi$_3$Al$_9$\cite{Ge:22}  and recently a new class of promising 4f (RE =Tm (J=6),Ho (J=8)) honeycomb materials BaRE$_2$(SiO$_4$)$_6$ was found \cite{liu:23}. All of these compounds have integer total angular momentum $J$ and thus will exhibit a CEF singlet ground state in this site symmetry.
}\\

{\BLU
\subsection{The case of P\lowercase{r}I\lowercase{r}$_3$ : a new near-critical singlet induced ferromagnet}
\label{sec:PrIr3}
}

{\BLU Now we consider a most recent addition to these examples in more detail, in the context of the previous theoretical analysis, which is the intermetallic compound \PR~described in Ref.~\onlinecite{gornicka:24}.}
It exhibits low temperature ferromagnetic order at $T_m\equiv T_C= 7.5\;\text{K}$. Because of the low site symmetry $C_2$ of Pr ions in the low temperature phase only singlet CEF states can be present.  {\BLU Due to its intermetallic nature the effective intersite interaction between
the Pr$^{3+}$ localised 4f moments is dominated by the RKKY mechanism mediated by conduction electrons \cite{jensen:91}.} Then the FM ordering should be of the induced moment type discussed here and previously \cite{thalmeier:24}. 
Indeed the observed anomaly of the specific heat (sofar only in zero field) around the ordering temperature supports such interpretation. It shows the two signatures of the induced moment magnet as evident from the experimental data in Fig.~\ref{fig:gacomp} adapted from Ref.~\onlinecite{gornicka:24}:
Firstly an underlying Schottky type anomaly resulting from the single-site CEF singlet-singlet splitting. Secondly the
specific heat jump associated with the induced ferromagnetic order. As explained before the relative position of the maximum $(T^{C_V}_{max})$ and the ordering temperature $T_m$ depends on the control parameter $\xi_s$. In the case of \PR~it was observed that the maximum $(T^{\gamma_V}_{max})$ of $\gamma_V=C_V(T)/T$  coincides with the ordering temperature. Then, following the arguments in the previous section this leads to values $\De=24.4\;\text{K}$ for the singlet splitting and a control parameter $\xi_s=1.08$ for \PR. Therefore the new compound \PR~may be viewed as a induced singlet ground state ferromagnet rather close to the QCP at $\xi_s=1$. In fact when plotting $C(T)$ itself for $\xi_s=1.08$ the specific heat jump at $T_m=7.5\;\text{K}$ lies already considerably below the Schottky maximum. For this asymptotic case with $\xi_s=1+\delta$ ($\delta\ll 1$) the relative jump size vanishes like \cite{thalmeier:24} $(C_V^- - C_V^+)/C_V^+\approx \de|ln\frac{\de}{2}|$ where $C_V^\pm$ are the values close above and below the transition \cite{thalmeier:24}.

{\BLU  We stress that we do not intend  a quantitative comparison of specific heat temperature dependences because \PR~ has a linear low temperature itinerant specific heat contribution (subtracted in Fig.~\ref{fig:gacomp})  and also presumably contributions from higher CEF states at elevated temperatures (the entropy exceeds $ln2$ there), both effects are not described by our two-singlet model. But the comparison in Fig.~\ref{fig:gacomp} demonstrates a similarity of shapes for $C_V/T$: A jump due to FM order sitting right on top of the Schottky peak.  The essential point we want to make is that the relative position $T_{max}$ of low temperature Schottky background maximum and ordering temperature T$_m$ gives information on the control parameter and the associated distance to the FM QCP of the induced moment magnetism (see also Fig.~\ref{fig:CVcont}).}

We finally note that the case where specific heat jump at $T_m$ is directly on top of the Schottky anomaly at $T_{max}$ of $C_V(T)$ itself occurs for $\xi_s\simeq 1.2$. For control parameters around this value  entropy release by ordering and level depopulation may be considered as comparatively important (Fig.~\ref{fig:relent}). This value of $\xi_s$ has mostly been adopted in the figures which are not directly related to \PR.\\

\section{Field dependence of magnetic exciton modes}
\label{sec:magex}

In the molecular field picture the inter-site exchange in the singlet-singlet system leads to
possible induced order with spontaneous moment creation due to singlet mixing and a concomitant
renormalisation of the CEF splitting of singlets, both depending on the external field as well.
Considering the effect of the exchange beyond this static picture, i.e., in an RPA response function 
technique \cite{jensen:91,thalmeier:24} allows to study the dynamics of the model. {\BLU An alternative for low temperatures
is provided by using the Bogoliubov transformation approach \cite{thalmeier:24}.} The finite CEF singlet gap
then turns into dispersive magnetic exciton modes due to the inter-site exchange which are already present in the disordered phase.
They have been investigated experimentally \cite{birgeneau:71,buyers:75,mcwhan:79,kawarazaki:95,savchenkov:19} as well as theoretically \cite{fulde:72,cooper:72,thalmeier:21,thalmeier:24} for numerous singlet ground state magnets. Their temperature dependence is characterized by (mostly arrested) soft mode behavior above
the ordering temperature
around the wave vector of the incipient induced magnetic order. As shown in detail in Ref.~\cite{thalmeier:24} the mode softening depends crucially on the nature of the first excited state, whether it is a singlet, doublet or triplet.

Here we focus on the two-singlet system for FM as well as AFM case but include the effect of the external field
on the evolution of magnetic exciton dispersion and in particular soft mode characteristics. These excitations may
be obtained from the dynamic version of Eq.~(\ref{eq:statsus0}) given by \cite{jensen:91,thalmeier:02,thalmeier:24}
\bea
\hspace{-0.5cm} 
\chi_{0\lam}(\om,T,h')&=&m_s^2\sum_{\tau\neq\tau'}|\la \tau|S_x|\tau'\ra|^2_\lam
\frac{p^\tau_\lam-p^{\tau'}_\lam}{\eps^{\tau'}_\lam-\eps^\tau_\lam-\om},
\label{eq:dynsus0}
\eea
where $\tau=\pm$ denote the mixed singlet eigenstates on sublattice $\lam=A,B$.
For finite frequency this contains only the vanVleck term without the zero-frequency pseudo-Curie term in the static case (Eq.(\ref{eq:statsus0})). It may be evaluated as
\bea
\chi_{0\lam}(\om)=\bigl(\frac{m_s^2}{2\hDe^\lam_T}\bigr)
\frac{\De}{(\De\hDe_T^\lam)^2-(\om)^2}f_s^\lam(T).
\label{eq:dynsus1}
\eea
Inserting this into the Eqs.(\ref{eq:statsuFM},\ref{eq:statsuAF}) according to $\chi_{0\lam}\rightarrow \chi_{0\lam}(\om)$ etc. leads to the dynamical collective RPA susceptibilities. The magnetic exciton  modes are then obtained from their poles for real frequencies, i.e. by the equation $I(\bq)\chi_{0}(\omega)=1$ for FM  and para AFM state and $I^2(\bq)\chi_{0\lam}(\omega)\chi_{0\bar{\lam}}(\omega)=1$ in the AFM ordered case. Explicitly the solution leads to the magnetic exciton dispersions given  by
\bea
&&\text{FM (para-ordered):}\non\\
&&\omega_F(\bq,T,h')=\De\hDe_T\bigl[1-\frac{\xi_s}{\hDe_T^3}f_s(T)\gamma(\bq)\bigr]^\fs,
\label{eq:dispFM}
\eea
\vspace{-0.2cm}
\bea
\label{eq:dispAF-para}
&&\text{AFM (para):}\non\\
&&\omega_0(\bq,T,h')=\De\hDe_T\bigl[1+\frac{\xi_s}{\hDe_T^3}f_s(T)\gamma(\bq)\bigr]^\fs,
\eea
\vspace{-0.2cm}
\bea
\label{eq:dispAF} 
&&\text{AFM-ordered:}\non\\
&&\omega_\pm(\bq,T,h')=\De\Bigl\{\fs(\hDe_T^{A2}+\hDe_T^{B2})\\
&&\pm\bigl[\frac{1}{4}(\hDe_T^{A2}-\hDe_T^{B2})^2+\xi_s^2\frac{f_s^A(T)f_s^B(T)}{\hDe_T^A\hDe_T^B}\ga(\bq)^2\bigr]^\fs
\Bigr\}^\fs .\non 
\eea
It is useful to consider a few instructive special cases for these mode dispersions in the paramagnetic as well as magnetically ordered temperature and field regimes. 

In the FM case Eq.~(\ref{eq:dispFM}) is appropriate
in the whole temperature and field  range with $\hDe_T$ obtained from Eq.~(\ref{eq:MF}) and $f_s(T)$ from  Eq.~(\ref{eq:aux}). For zero field a true phase transition occurs for $\xi_s>1$. For $T\rightarrow T_m$ we have $\hDe_{T}\rightarrow 1$ and then $f_s(T)\rightarrow \xi_s^{-1}$. Furthermore at the FM ordering vector $\bq=0$ we get $\ga(0)=1$. Then Eq.~(\ref{eq:dispFM}) leads to $\omega(0,T_m,0) \rightarrow 0$, i.e. a soft mode at the FM $\Gamma$-point appears as a precursor to the induced FM ordering. For finite field when only a crossover to a FM polarized state occurs the soft mode acquires a finite frequency again already at small field sizes.

The soft mode precursor behaves even more interesting in the AFM phase transition. For $T<T_m(h')$ there are now generally two modes from Eq.~(\ref{eq:dispAF}) because $\hDe_T^A\neq\hDe_T^B$ on the two sublattices. In the paramagnetic phase with $\hDe_T^\lam=\hDe_T$ it reduces to the same form
as Eqs.~(\ref{eq:dispFM},\ref{eq:dispAF-para}) but with $\omega_\pm(\bq)$  corresponding to a downfolding  from the simple cubic paramagnetic Brillouin zone (BZ) to the fcc antiferromagnetic BZ (AFBZ). The $1.$ AFBZ is defined by the boundaries $\bq\cdot\bQ'=\pm1$ where $\bQ'$ is any of the eight midpoint (L) vectors $(\pm\fs,\pm\fs,\pm\fs)$. The unfolding in the paramagnetic phase is defined by $\omega_+(\bq)=\omega_F(\bq)$ and  $\omega_-(\bq)=\omega_0(\bq)$ in the $2.$ AFBZ and vice versa in the $1.$ AFBZ (the complement in the simple cubic BZ) as further discussed in Sec.~\ref{sec:discussion}.

\begin{figure}
\includegraphics[width=0.80\columnwidth]{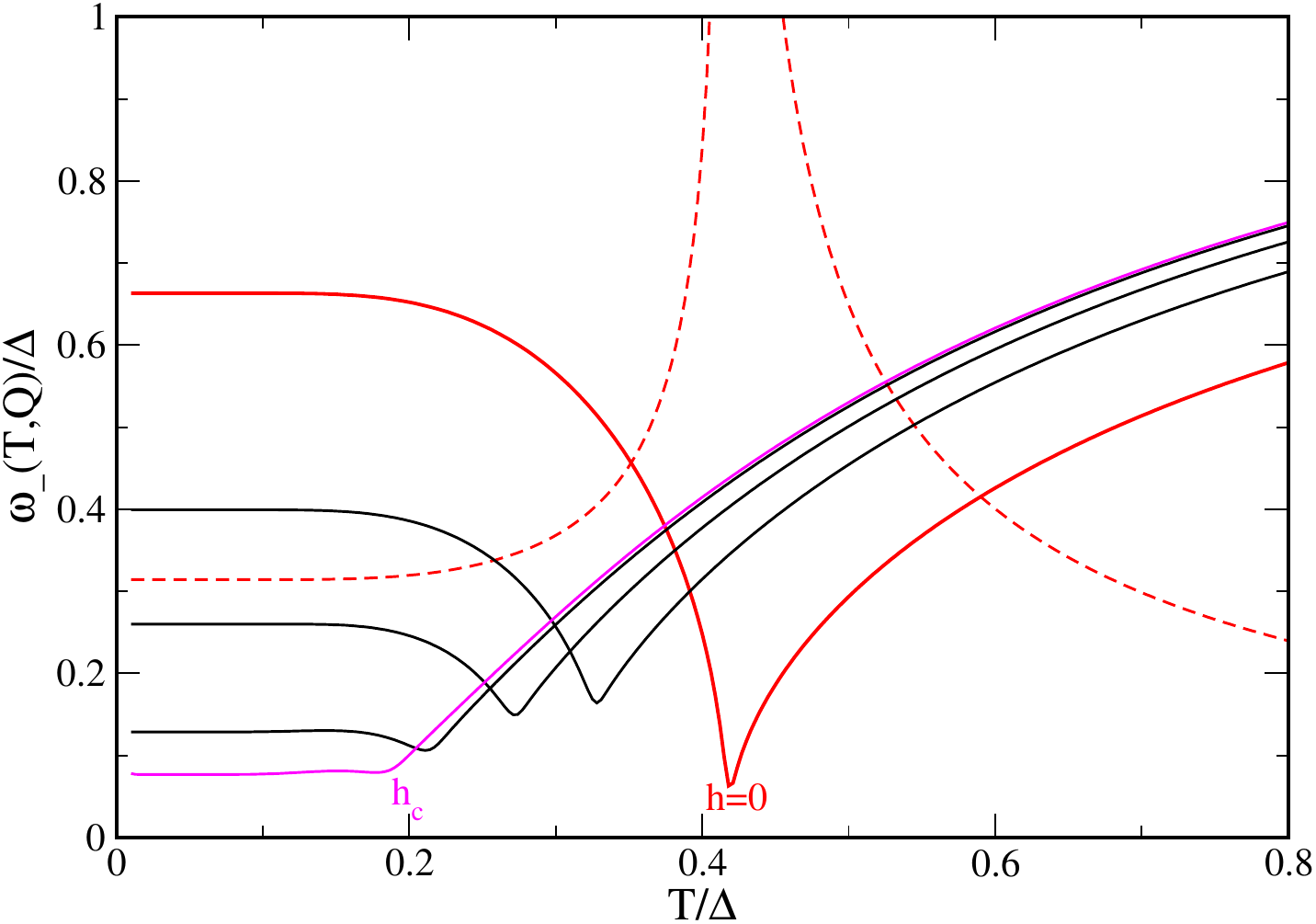}
\caption{Temperature dependence of the AFM arrested soft mode $\omega_-(T,\bQ)$ at various fields 
$h/\De=0, 0.6,0.7,0.75\approx h_c$ from right to left. Between $h=0$ and $h_c$ the critical mode for induced order
acquires a finite frequency again (cf. Fig.\ref{fig:AF-Tm}(a)). The dashed red line is the temperature dependent intensity
(Eq.~(\ref{eq:AFMspec})) of the $h=0$ soft mode.
}
\label{fig:softmode}
\end{figure}

With similar arguments as above for $\xi_s>1$ and zero field the mode frequencies at the AFM ordering vector $\bQ=(1,1,1)$ (R-point) are $\omega_\pm(\bQ,T_m,0)=0, 2\De$ with $\omega_-(\bQ)$ now corresponding to the precursor soft mode. In the AFM case the transition at $T_m(h')$ exists also for finite fields $h'<h'_c$ below the critical one. Then the question arises whether the zero-field soft mode appearance at $T_m$ seen before persists also for finite field for all  $T_m(h')$. This is, however, not the case for the following reason: The AFM critical temperature as determined by Eq.~(\ref{eq:tmh}) contains
also the effect of the (second) zero-frequency pseudo-Curie terms leading to a slightly larger $T_m(h)$ than to be expected from the (first) vanVleck terms only, However the mode frequency is determined solely by the poles due to the dynamical vanVleck terms and therefore the AFM transition $T_m(h')$ happens already before the complete softening of the precursor mode takes place. This 'arrested soft mode' frequency is obtained from Eq.~(\ref{eq:dispAF}) as
\bea
\label{eq:softmode}
&&\omega_-(\bQ,T_m(h'),h')=\non\\[0.3cm]
&&\De\bigl(\frac{\xi_s\De}{2T_m}\bigr)^\fs\bigl(\hDe^2_{T_m}-1\bigr)^\fs(1-f_s^2(T_m))^\fs\geq0,
\eea
which vanishes in the zero field limit where $\hDe_{T_m}=1$ and also at the critical field where $T_m(h'_c)\rightarrow 0$ leading to $f_s(T_m)\rightarrow 1$.  Therefore strictly, even in a singlet-singlet model a true soft mode occurs only at $T_m(h')$ for $h'=0,h_c'$ while an 'arrested' soft mode with $\omega_-(\bQ,T_m(h'),h') >0$ is observed in the interval $0<h'<h'_c$. (In a singlet-triplet induced moment magnet this happens already  at zero field due to the effect of the genuine Curie terms in the static susceptibility resulting from the excited triplet \cite{smith:72,thalmeier:24}). The naively unexpected behavior of $\omega_-(\bQ,T_m(h'),h')$ is clearly seen in Figs.~\ref{fig:AF-Tm}(a) and Fig.~\ref{fig:softmode}.
%
\begin{figure}
\includegraphics[width=0.99\columnwidth]{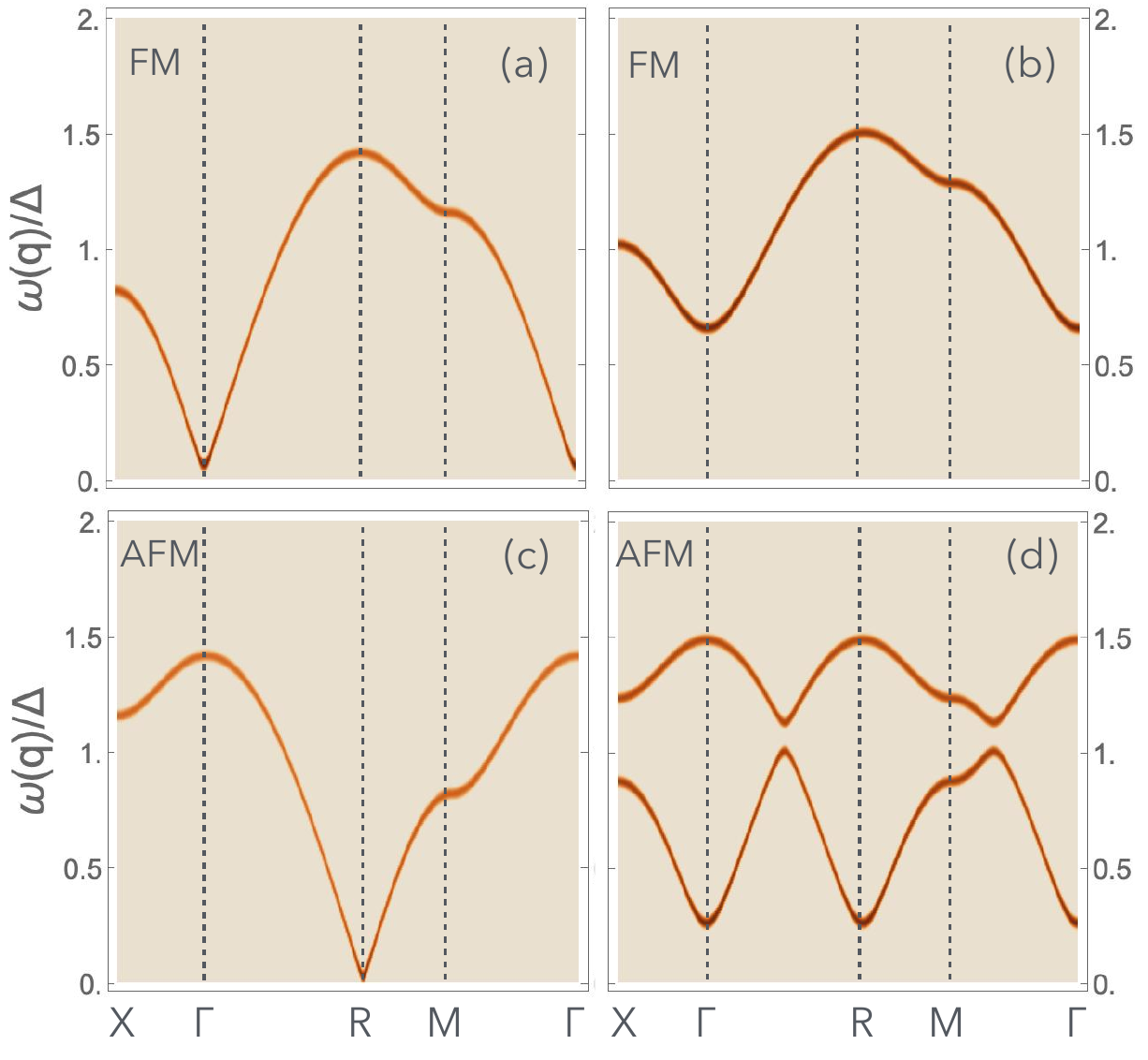}
\caption{Magnetic exciton dispersions for $\xi_s=1.2$ in the simple cubic BZ along  $X(010), 
\Gamma (000),  R(111), M(110), \Gamma(000)$ path. This is an extended zone scheme with respect to the two-sublattice
AFM (d) ordered phase.  (a) FM zero-field critical case at the transition temperature $T_m/\De=0.417$ with soft mode at $\Gamma$ point. (b) FM again at $T_m$ for finite field $h/\De=0.05$ with rapid gapping of the soft mode at $\Gamma$.
(c) AFM zero-field critical case at the transition temperature $T_m/\De=0.417$ with soft mode at R(111) point.
(d) AFM at finite field ($h/\De$=0.6 and $T/\De$ =0.3 ) inside the ordered phase. The soft mode will be gapped and hybridization of upper and lower modes occur (in the extended zone scheme of the AFM R and $\Gamma$ are equivalent and the L$(\fs,\fs,\fs)$ midpoint of $\Gamma$-R segment is an AFBZ zone boundary point). For soft mode field dependence along $T_m(h)$ transition line see Fig.~\ref{fig:AF-Tm}.  A logarithmic scale for better visibility of dispersions has been employed for the mode intensity.
}
\label{fig:disp}
\end{figure}
%

To complement the dispersions we also calculate the intensities of the modes, i.e. the spectral function (without the Bose prefactor) which is proportional to the inelastic neutron scattering cross section. It is  obtained from
\bea
S(\bq,\omega)=\frac{1}{\pi}Im\sum_\lam\hat{\chi}_{\lam\lam}(\bq,\om)_{\om\rightarrow \omega+i\eta}.
\eea
Again the sum over sublattices is absent for FM and paramagnetic AFM cases.
Using the dynamical form of the RPA-susceptibilities in Eq.~(\ref{eq:statsuFM},\ref{eq:statsuAF}) and employing Eq.~(\ref{eq:dynsus1}) we get for the single FM mode dispersion (Eq.~(\ref{eq:dispFM})):
\bea
FM: \;\; S(\bq,\omega)=\frac{1}{2}m_s^2\frac{\hf_s(T)}{2\home_\bq}\delta(\omega-\omega_\bq),
\label{eq:FMspec}
\eea
with the definition $\hf_s(T)=f_s(T)/\hDe_T$ implied (cf. Eq.(\ref{eq:aux})) and $\home_\bq=\omega_\bq/\De$. Likewise the spectral function for the two AFM dispersive modes (Eq.~(\ref{eq:dispAF})) is given by
\bea
AFM:\;\; S(\bq,\omega)&=&\frac{1}{2}m_s^2\sum_{n=\pm}\frac{\hf_{sn}(T)}{2\home_{n\bq}}\delta(\omega-\omega_{n\bq})\non\\[0.3cm]
\hf_{sn}(T)&=&\frac{\sum_\lam\hf_s^\lam(T)(\home^2_{n\bq}-\hDe^{\blam 2}_T)}
{(\home^{n 2}_\bq-\home^{\bn 2}_\bq)},
\label{eq:AFMspec}
\eea
where $n=\pm$; $(\bn=\mp)$ denotes the two modes and $\lam=A,B$; $(\blam=B,A)$ is as before the sublattice index.
Furthermore as above we introduced  $\hf_s^\lam(T)=f_s^\lam(T)/\hDe^\lam_T$ (cf. Eq.(\ref{eq:auxlam})) and $\home_{n\bq}=\omega_\bq/\De$. It is obvious from the FM case but also true for AFM that the mode intensity increases with decreasing frequency 
and for the soft mode diverges (Fig.\ref{fig:softmode}). The mode dispersions and spectral functions for some typical and instructive
choices of field and temperature in ordered and paramagnetic phases and their implications are explained below.
%
%

\section{Discussion of numerical results}
\label{sec:discussion}

The basis for understanding the physical effects in the two singlet model  is the determination of the induced oder parameters  given by the solutions of the selfconsistent solutions of Eq.~(\ref{eq:MF}). Two examples $(\xi_s=1.2)$ for FM (temperature dependence)  and AFM (field dependence) are shown in Fig.~\ref{fig:FM-AFOP}. For the former the zero-field FM transition immediately turns into a crossover behaviour at finite field since the appearance of a magnetisation $ \la S_x\ra$ is no longer a symmetry breaking in external fields. The same is true for the renormalised singlet splitting $\hDe_T$.
The AF case shown in (b) demonstrates the two sublattice order parameters $\la S_x\ra_{A,B}$ as well as associated staggered and homogeneous magnetisation as function of field up to the critical one $h_c(T)$. The corresponding field dependence of renormalised splittings $\De^\lam_T(h)$ for A,B  sublattices is non-monotonic as function of field, in particular for B sublattice which is originally polarised opposite to the external field. This is quite distinct from the FM case and has consequences for the magnetocaloric effects as discussed below.\\

From the solution of Eq.~(\ref{eq:MF}) we can also determine the phase boundary of the induced moment AFM in the h-T plane. It is shown in Fig.~\ref{fig:AF-Tm}(a) for different $\xi_s$. The $T_m(h)$ dependence close to the critical field $h_c$ is remarkably steep exhibiting a logarithmic singularity. This is in contrast to common AFM with degenerate ground state where it has a square root singularity. The concomitant dependence of the arrested soft mode frequency $\omega_-(\bQ,T_m(h)$ (Eq.~(\ref{eq:softmode})) when $(T,h)$ are varied along the phase boundary is also shown by dashed lines in Fig.\ref{fig:AF-Tm}(a). It demonstrates that a true soft mode is only realised at the boundary points $h=0,h_c$ but not in between as further discussed below. In Fig.~\ref{fig:AF-Tm}(b) we also present the dependence of zero field transition temperature $T_m(\xi_s)$ and critical field $h_c(\xi_s)$ as function of control parameter above the QCP $(\xi_s=1)$. The asymptotics for $\xi_s\rightarrow 1$ is given by different type of singular behaviour according to Eq.~(\ref{eq:critasymp}).\\

The homogeneous susceptibility anomaly is a primary method to characterise H-T phase diagrams by tracing their anomalies. In Fig.~\ref{fig:FMAF-susz-t} we present the results for both cases. Since only $J_z$ operator has nonzero matrix elements between the singlet only a longitudinal (zz) susceptibility has to be considered. For FM the singularity at $T_m$ that signifies the order is rapidly suppressed in an external field. As shown below the suppression occurs in parallel with that of the specific heat anomaly at the transition. In the AFM the \bq = 0 susceptibility stays finite at the transition with a cusp anomaly appearing that shifts with $T_m(h)$ to lower values (c.f. Fig.~\ref{fig:AF-Tm}(a)) until it vanishes at the critical field $h_c$. For higher field a plateau due to the pure vanVleck term is present which is shifted downward because of the increasing singlet splitting. An important point is the
depression $\delta\chi_{AFM}(h)$ (Eq.~(\ref{eq:dropsus})) from peak height at $T_m(h)$ to zero temperature. In degenerate ground state moments
the longitudinal susceptibility drops to zero for the fully polarised AFM at T=0. This is {\it not} the case in an induced moment magnet
because the saturation moment (in units of $g_J\mu_B$) is $\frac{m_s}{2\xi_s}(\xi_s^2-1)^\fs$ which depends on $\xi_s$ and is smaller than the maximum value $\fs m_s$, reaching zero at the QCP $\xi_s$=1. Therefore it  still can be polarised by a probe field and hence the longitudinal susceptibility is nonvanishing at $T=0$. At zero field the depression $\delta\chi_{AFM}(0)$ is given by Eq.~(\ref{eq:dropsus}) and only on approaching the degenerate case for $\xi_s\gg 1$ it reaches maximum value one, i.e. $\chi(0)$ vanishes. For fixed $\xi_s$ and increasing field the opposite
is observed: The depression becomes less and vanishes at the critical field as summarised in the inset of Fig.~\ref{fig:FMAF-susz-t}.
This variable depression  $\delta\chi_{AFM}(h;\xi_s)$ of the longitudinal susceptibility{\BLU~which depends }on field as well as control parameter is characteristic for the induced moment magnets and might be used to quantify $\xi_s$ from experiment in the AFM case.\\

 {\BLU Before discussing the specific heat it is useful to consider an important symmetry aspect
 of the induced magnetic order. For FM only time reversal
symmetry is broken while for AFM both time reversal and translational symmetry are broken (although
the product is still conserved).
Application of an external field removes time reversal symmetry and then in the FM there is no
more symmetry breaking because the homogeneous magnetisation $M_x^0$ (Eq.(\ref{eq:mag})), which is
the order parameter at zero field is now already present above $T_m(h=0)$. Therefore at finite
field no phase transition but only a gradual crossover to a more polarised state will appear around and below this temperature.
This leads to the rapid suppression of the zero-field specific heat jump in finite field.
On the other hand for AFM the translational symmetry breaking due to the two sublattices is still present
in finite field, characterised by the staggered order parameter $M_x^s$ (Eq.(\ref{eq:mag})). Consequently
the induced AFM phase transition survives up to the critical field as long as  $M_x^s$ is finite (Fig.~\ref{fig:FM-AFOP}(b)).
These different symmetry properties of the two types of order will be reflected in their strikingly distinct field dependence
of the specific heat.}

The specific heat evolution of the singlet-singlet magnet as function of field and temperature is shown
in Figs.~\ref{fig:AFFM-cv-t}(a,b) for AFM and FM, respectively. 
In both figures the dashed black curve corresponds to the background Schottky anomaly $C_V^0(T)$ resulting from the noninteracting split singlet-singlet CEF states and the full black curve corresponds to the interacting  zero field case  with $\xi_s=1.2$. It is valid for both FM and AFM transition at zero field. For this control parameter the specific heat jump of
the zero field transition $(T_m)$ sits right on top of the background Schottky anomaly peak $(T_{max})$.
Then for AFM case in Fig.~\ref{fig:AFFM-cv-t}(a) below $T_m$ the entropy release is due to the ordering as well as due to the background depopulation of the excited singlet as further discussed below.

 For finite field the FM transition in  Fig.~\ref{fig:AFFM-cv-t}(b) at $T_m$ rapidly turns into a crossover and the transition peak merges with the background. On the other hand for the AFM case the ordering at $T_m(h)$ prevails at finite field but shifts to lower temperature (see also Fig.~\ref{fig:AF-Tm}(a)) with the specific heat jump sliding down on the left flank of the background Schottky anomaly and being continuously reduced in size. The $\delta C_V(T_m(h))$ jump normalised to the maximum of $C_V^0(T_{max})=0.44 k_B/\text{site}$ is shown in the inset as function of field up to the critical  $h_c$ for this control parameter where it vanishes. 
 
 As a specific example of the competing presence of Schottky anomaly and ordering specific heat jump we have discussed the case of ferromagnetic \PR~in Sec.~\ref{sec:PrIr3} as presented in Fig.~\ref{fig:gacomp} where we showed that from the relative position of the two peaks one may estimate the control parameter as $\xi_s\approx 1.08$ in this case, which is already rather close to the QCP of $\xi^c_s=1$ below which magnetic order is suppressed. {\BLU The field-dependence of the magnetic specific heat of \PR~which should look qualitatively similar to Fig.~\ref{fig:AFFM-cv-t}(b)
 has not yet been determined experimentally.}
 \\.
 
The magnetocaloric properties of the singlet-singlet model are also presented by the contour plots of $C_V$ in Fig.~\ref{fig:CVcont} in the field-temperature plane. The comparison of  FM and AFM cases shows their  distinct appearance since the magnetic phase transition survives in the latter up to the critical field while in the former it is rapidly
suppressed and turned into a crossover. The latter is seen in Fig.~\ref{fig:CVcont}(a),(c) for two different  control parameters $\xi_s$. For finite field the polarisation $\la S_x\ra$ and hence the singlet effective splitting $\hDe_T(h)$ increase rapidly with field shifting the Schottky anomaly to higher temperature. The AFM case shows more intricate behaviour in Fig.~\ref{fig:CVcont}(b),(d): The contour of the phase boundary $T_m(h)$ is clearly visible but becomes blurred at lower $T_m$ on approaching $h_c$ because the specific heat jump is strongly suppressed in this region (see inset of Fig.~\ref{fig:AFFM-cv-t}(a)). For $\xi=1.1$ the phase boundary
$T_m(h)$ and broad Schottky maximum at $T_{max}(h)$ are separated while for $\xi_s=1.2$ they coincide. The latter shows now much less field dependence because there are two singlet excitation energies $\hDe^\lam_T(h)$ at the two sublattices $\lam=A,B$. One of then (A) increases and the other (B) decreases (because the polarisation is opposite to the external field, see Fig.\ref{fig:FM-AFOP}(b)). Then the thermal averaging leads to a weak field dependence of the background Schottky anomaly in this case for fields up to the critical field $h_c$.

We emphasise that the two coexisting channels of entropy release, depopulation and ordering are characteristic for the induced moment magnetism. In a degenerate ground state magnet only the latter is active. We show this explicitly in
Fig.~\ref{fig:relent} which presents the comparison of total entropy $S(T)$ and background single-site entropy $S_0(T)$
in the upper panel. The right panel presents the relative entropy from ordering compared to the one from depopulation, ie.,
the ratio $(S_0(T)-S(T))/(\ln 2-S_0(T))$ where $\ln 2$ is the high temperature limit of S(T). It demonstrates that it increases strongly with $\xi_s$ when the release from ordering becomes dominant and finally exclusive in the limit $\xi_s\gg1$ where
the level splitting will be irrelevant.

The field dependence of specific heat also provides the basis for the adiabatic magnetocaloric cooling rate $\Gamma_{mc}(T,h)$ as given by Eq.~(\ref{eq:mce}) and shown for the FM and AFM case in Fig.~\ref{fig:mce}.
In the former, when there is no phase transition at finite field it behaves smoothly, increasing at lower field and decreasing
at lower temperature mostly due to the behaviour of the temperature derivative of the magnetisation in Fig.~\ref{fig:mce}.
For $h\rightarrow 0$ it diverges when $T\rightarrow T_m^-$ on approaching the FM transition. This is because $M_x^0$ becomes
the FM order parameter in that limit with a diverging temperature derivative. On the other hand for $T>T_m$ when $M_x^0$ is no longer singular for small fields $\Gamma_{mc}$ will drop to zero. For larger fields it will fall off $\sim 1/h$ in agreement with the noninteracting case (Appendix \ref{app:freecase}).

For the AFM case pronounced anomalies and a sign change of $\Gamma_{mc}$ are seen at the critical field $h_c$. In the ordered
phase as well in the para phase ($T>T_m$)  it approaches zero at small fields since the temperature derivative of  $M_x^0$ always tends to zero in that limit for the AFM. For fields $h\gg h_c$ and  much larger than in the FM case  $\Gamma_{mc}$ will achieve a maximum and then fall off like $1/h$ as in the noninteracting limit. The maximum field in Fig.~\ref{fig:mce}(b) is still below that range.
The integrated adiabatic temperature change when decreasing the field is negative at all fields for FM while it is negative
above $h_c$ and slightly positive below for the AF case.

Instead of sweeping the field one may apply rapid adiabatic pressure change leading to the analogous barocaloric cooling rate  $\Gamma_{bc}(p)$ at a constant applied field (Eq.~(\ref{eq:bce})). It is shown in Fig.~\ref{fig:bce} as function of pressure for different  fields and fixed temperature. For zero field the FM and AFM pressure dependence of $\Gamma_{bc}$ is identical. The exhibit a sign change and a jump at the critical pressure above which magnetic order vanishes (due to the decrease of $\xi_s(p)$). For finite fields they are largely different. While for the FM  the barocaloric coefficient develops a smooth
pressure dependence in the AFM ordered regime the step prevails and moves to lower critical pressures as field values increase up to the critical one where it vanishes. {\BLU If the field is kept constant in both FM and AFM cases but temperature is increased, then in each of the curves in  Fig.~\ref{fig:bce} the step anomaly (broadened for FM) is shifted to lower pressure, but the shape of the $\Gamma_{bc}(p)$  curve is qualitatively preserved.}\\

Finally we discuss the field and temperature dependence of the singlet-singlet magnetic exciton spectrum as presented
in the panels in Fig.~\ref{fig:disp}. 
For FM as well as AFM mode presentation we use the paramagnetic simple cubic BZ. It is an extended zone scheme for the AFM. In this case the $\Gamma$- and R-points are equivalent points in the AFM phase connected by $\bQ=(111)$ ordering vector and correspond both to centers of the fcc AFBZ. Their midpoint L($\fs,\fs,\fs)$) is the AFBZ zone boundary point.\\
In the FM case the mode frequency $\omega_F(\bq)$ (Eq.~(\ref{eq:dispFM})) at the 
$\Gamma$ point softens at the zero-field transition temperature T$_m$ as seen in Fig.~\ref{fig:disp}(a).  For already quite small fields a re-hardening with a gap appearing at $\Gamma$ is observed. The same is true in the zero field case
when temperature is lower than $T_m$ inside the FM phase.\\
The more interesting AFM case is presented in the lower panels. Fig.~\ref{fig:disp}(c) shows the magnetic exciton dispersion (Eq.~(\ref{eq:dispAF-para})) of the AFM right at the transition temperature where the lower mode $\omega_0(\bQ)$  becomes soft at the AFM ordering vector (R-point). For larger field but inside the AFM phase two modes appear due to the downfolding to AFBZ accompanied by gapping and hybridization at the new AFBZ zone boundaries. This is due to the fact that $\hDe_T^A\neq\hDe_T^B$ for the two sublattices. The temperature dependence of the lower mode for fields increasing from zero to $h_c$ is shown in Fig.~\ref{fig:softmode}. The softening in the para phase above $T_m(h)$ is clearly seen but the 
soft mode appears only at $h=0,h_c$. 
For fields inside the interval
$(0,h_c)$ the soft mode is arrested due to the subtle effect of the pseudo-Curie terms on the the transition temperature as explained before in Sec.~\ref{sec:magex}. The resulting arc-like field dependence of $\omega_-(\bQ,T_m(h),h)$ visible in Fig.~\ref{fig:softmode} is even more clearly seen  in Fig.~\ref{fig:AF-Tm}(a) based on the analytical formula in Eq.~(\ref{eq:softmode}).

\section{Summary and Conclusion}
\label{sec:summary}

In this work we have studied primarily the magnetocalorics and  field-dependence of magnetic exciton modes of a simple singlet-singlet model. It describes schematically  the magnetism in numerous 4f and 5f compounds with integer occupied f shells (non-Kramers ions) in a radically simplified manner: All excited CEF states are assumed to be at sufficiently  high 
energies as compared to the transition temperature and the lowest CEF singlet-singlet splitting. Since both
singlet states are nonmagnetic the magnetism in this model can only occur by a spontaneous admixture of
ground- and excited states generating an induced moment.

 This genuine quantum magnetism shows quite
different characteristics as compared to the quasiclassical magnetic order associated with degenerate magnetic ground
state (Kramers ions). Firstly the order occurs only if a control parameter involving the ratio of intersite exchange
and level splitting exceeds a critical value $(\xi_s=1)$ which defines a quantum critical point separating the paramagnetic
and magnetic ground state. Both transition temperature and critical field show singular dependence on the control parameter $\xi_s$ above the QCP.

The specific heat appearance is connected with the possible low temperature entropy release shows two contributions:
Firstly from the upper level depopulation leading to the Schottky-type background anomaly with a maximum at $T_{max}$
and secondly a specific heat jump at the critical temperature $T_m$ for induced order. Their relative position is determined by the control parameter. For $\xi_s\approx 1.2$ the two anomalies are on top of each other and both give a significant contribution to the entropy release. The recently found induced moment singlet-singlet FM compound \PR~is rather close to that situation.

In external field only the AFM transition can survive whereas the FM changes to crossover behaviour. In the former case
the specific heat jump at $T_m(h)$ is shifted to lower temperature sliding on the lower flank of the almost field independent Schottky anomaly and reducing is size until it vanishes at the critical field. This means that for a control parameter where the zero field 
$T_m$ and $T_{max}$ are close they will rapidly be suppressed in an applied field. This should be a distinct characteristic of induced moment AFM. On the other hand for the FM the specific heat jump at zero field
is quickly absorbed in the Schottky anomaly that shifts to higher temperatures with increasing field.

The rather different magnetocaloric characteristics of induced singlet  FM and AFM are also reflected in the adiabatic
cooling rate. For FM it is continuously increasing with lowering field or increasing temperature [for T$<$T$_m(0)$]. On the other hand for the AFM it changes sign at $T_m(h)$ and is strongly reduced further below reaching zero at low fields. And a similar distinction is observed in the barocaloric coefficient.

The critical properties of the induced moment magnets are also reflected in the excitation spectrum, i.e., the magnetic
exciton dispersions and their field and temperature dependence within and outside the ordered region.
In FM and AFM zero field case a soft mode appears at the respective ordering wave
vectors $\bq =0,\bQ$ at $T_m$ and a re-hardening of the mode occurs below the transition. 
In the AFM case a surprising field dependence
is observed: When moving along the phase boundary $T_m(h)$, a true soft mode is found only at the endpoints 
$h=0,h_c$, in between an  arrested soft mode with finite frequency is observed. This was shown to 
be due to induced pseudo-Curie term in the static susceptibility that lead to a $T_m(h)$ larger than the temperature
where complete softening could be achieved.

The numerous specific magnetocaloric and field- dependent magnetic exciton dispersive characteristics for the singlet induced moment magnets that have been derived in this work should be helpful to characterise localised f-electron magnets that may be approximately correspond to this class, depending on the concrete CEF level scheme of the compound.\\

\begin{acknowledgments}
The author thanks Burkhard Schmidt for technical assistance and discussions.
\end{acknowledgments}

\appendix

\section{Adiabatic caloric coefficients for noninteracting two singlet model}
\label{app:freecase}

As a reference to the FM and AFM models it is useful to derive the adiabatic caloric coefficients
for the case of isolated two-level systems. Using their definitions in Sec. ~\ref{sec:calcoeff} and setting
$I_e$= 0 or $\xi_s$= 0 in the thermodynamic potentials we can derive the following expressions (for clarity of dimensions
we explicitly include k$_B$):
\bea
\Gamma^0_{mc}(T,H)=\bigl(\frac{k_BT}{\De_h}\bigr)\bigl(\frac{m_sh}{\De_h}\bigr)\;\;\bigl[\frac{g_{eff}\mu_B}{k_B}\bigr],
\label{eq:mc0}
\eea
where $\Delta_h$=$[\De^2+(m_sh)^2]^\fs$ is the two singlet splitting in an external field but without interaction effect.
This expression  is proportional to temperature as well as field strength. This is quite different from the result for a degenerate
magnetic Kramers doublet that is, following a similar procedure,  given by
\bea
\Gamma^0_{mc}(T,H)=\bigl(\frac{k_BT}{m_sh}\bigr)\;\;\bigl[\frac{g_{eff}\mu_B}{k_B}\bigr] =\frac{T}{H},
\label{eq:mc0_deg}
\eea
which is inversely proportional to the field. The difference is due to the fact that in the degenerate case
the only energy scale present is the Zeeman splitting itself. Indeed, sending $\De\rightarrow 0$ in $\De_h$ 
the Eq.~(\ref{eq:mc0}) reproduces Eq.(\ref{eq:mc0_deg}).

Likewise we may derive the barocaloric coefficient for the isolated two-singlet system and obtain,
keeping the field at constant value:
\bea
\Gamma^0_{bc}(T,p)=\Bigl(\frac{k_BT}{\De(p)}\Bigr)\Bigl(\frac{\De(p)}{\De_h(p)}\Bigr)^3 k_B^{-1},\\\non
\label{eq:bc0}
\eea
with  $\Delta_h(p)$=$[\De(p)^2+(m_sh)^2]^\fs$.
Here we implied that the pressure in $(\partial T/\partial p)_S$ is given in units of $[c_B/\Omega_\De]$.
In these units the pressure dependence of the bare singlet splitting is given by $\De(p)=\De(1+p)$ in the 
Gr\"uneisen model. Without applied field this reduces to 
\bea
\Gamma^0_{bc}(T,p)=\Bigl(\frac{k_BT}{\De(p)}\Bigr)k_B^{-1},
\eea
which is formally the same as Eq.~(\ref{eq:mc0_deg}), but the pressure enters only implicitly through the
$\De(p)$ function which tends to the finite value $\De$ for $p\rightarrow 0$.

\bibliography{References}
\end{document}